\documentclass[12pt]{article}
\pdfoutput=1

\usepackage{amsmath}
\usepackage{amssymb}
\usepackage{epsfig}
\usepackage{graphicx}
\usepackage{cite}
\usepackage{multirow}
\usepackage{longtable}
\usepackage{lscape}
\usepackage{bbm}
\usepackage{dcolumn}
\usepackage{bm}
\usepackage{booktabs}
\usepackage{rotating}
\usepackage{mathrsfs}
\usepackage{feynmf}
\usepackage{slashed}

\allowdisplaybreaks[1] 
\def\lsim{\raise0.3ex\hbox{$\;<$\kern-0.75em\raise-1.1ex\hbox{$\sim\;$}}}

\newcommand{\capdef}{}
\newcommand{\mycaption}[2][\capdef]{\renewcommand{\capdef}{#2}%
        \caption[#1]{{\footnotesize #2}}}
\makeatletter
\renewcommand{\fnum@table}{\textbf{\tablename~\thetable}}
\renewcommand{\fnum@figure}{\textbf{\figurename~\thefigure}}
\makeatother

\newcounter{myenumi}

\renewcommand{\themyenumi}{\roman{myenumi}}
{\end{list}}

\newlength{\myem}
\newcounter{mysubequation}[equation]

\makeatletter
\renewcommand{\section}{\@startsection{section}{1}{0em}{-\baselineskip}%
{\baselineskip}{\normalfont\large\bfseries}}
\renewcommand{\subsection}%
{\@startsection{subsection}{2}{0em}{-0.7\baselineskip}%
{0.7\baselineskip}{\normalfont\bfseries}}
\makeatother

\newcommand{\bi}{\begin{itemize}}
\newcommand{\ei}{\end{itemize}}

\newcommand{\be}{\begin{equation}}
\newcommand{\ee}{\end{equation}}
\newcommand{\bea}{\begin{eqnarray}}
\newcommand{\eea}{\end{eqnarray}}

\newcommand{\trm}[1]{\textrm{#1}}
\newcommand{\mcl}[1]{\mathcal{#1}}

\newcommand{\ie}{{\it i.e.}}

\newcommand{\eg}{{\it e.g.}}

\newcommand{\cf}{{\it cf.}}

\newcommand{\eq}{Eq.}

\newcommand{\fig}{Fig.}

\newcommand{\Ref}{Ref.}
\newcommand{\Refs}{Refs.}
\newcommand{\Sec}{Sec.}

\newcommand{\App}{Appendix}

\newcommand{\Tab}{Table}

\newcommand{\equ}[1]{\eq~(\ref{equ:#1})}
\newcommand{\figu}[1]{\fig~\ref{fig:#1}}

\newcommand{\T}{\mathsf{T}}
\newcommand{\hc}{\mathrm{h.c.}}
\newcommand{\ii}{\mathrm{i}}

\newcommand{\ex}[1]{\cdot 10 ^{#1}}

\newcommand{\vectwo}[2]{\begin{pmatrix} #1 \\ #2 \end{pmatrix}}

\newcommand{\matrixtwo}[4]{\begin{pmatrix} #1 & #2 \\ #3 & #4 \end{pmatrix}}

\begin{document}

\begin{titlepage}

\renewcommand{\thefootnote}{\alph{footnote}}



\renewcommand{\thefootnote}{\fnsymbol{footnote}}
\setcounter{footnote}{-1}

\begin{flushright}
IFIC/11-44, MPP-2011-105
\end{flushright}

{\begin{center}
{\large\bf

Neutrino mass from higher than d=5 effective operators in SUSY, 
and its test at the LHC \\[0.3cm]

} \end{center}}
\renewcommand{\thefootnote}{\alph{footnote}}

\vspace*{.8cm}
\vspace*{.3cm}
{\begin{center} {\large{\sc
                Martin~B.~Krauss\footnote[1]{\makebox[1.cm]{Email:}
                martin.krauss@physik.uni-wuerzburg.de},
 		Toshihiko~Ota\footnote[2]{\makebox[1.cm]{Email:}
                toshi@mppmu.mpg.de}, \\
 		Werner~Porod\footnote[3]{\makebox[1.cm]{Email:}
                porod@physik.uni-wuerzburg.de}, and
                Walter~Winter\footnote[4]{\makebox[1.cm]{Email:}
                winter@physik.uni-wuerzburg.de}
                }}
\end{center}}
\vspace*{0cm}
{\it
\begin{center}

\footnotemark[1]${}^,$\footnotemark[3]${}^,$\footnotemark[4]
       Institut f{\"u}r Theoretische Physik und Astrophysik, Universit{\"a}t W{\"u}rzburg, \\
       Am Hubland, 97074 W{\"u}rzburg, Germany 
       
\footnotemark[2]
	Max-Planck-Institut f{\"u}r Physik (Werner-Heisenberg-Institut), \\
        F{\"o}hringer Ring 6, 80805 M{\"u}nchen, Germany

\footnotemark[3]
    AHEP Group, Instituto de F\'{\i}sica Corpuscular \\
    C.S.I.C./Universitat de Val{\`e}ncia \\
  E--46071 Val{\`e}ncia, Spain 

\end{center}}

\vspace*{1.5cm}

{\Large \bf
\begin{center} Abstract \end{center}  }

We discuss neutrino masses from higher than $d=5$ effective operators in a supersymmetric framework, where we explicitly demonstrate which operators could be the leading contribution to neutrino mass in the MSSM and NMSSM.
As an example, we focus on the $d=7$ operator $L L H_u H_u H_d H_u$, for which we systematically derive all tree-level decompositions. We argue that many of these lead to a linear or inverse see-saw scenario with two extra neutral fermions, where the lepton number violating term is naturally suppressed by a heavy mass scale when the extra mediators are integrated out. We choose one example, for which we discuss possible implementations of the neutrino flavor structure. In addition, we show that the heavy mediators, in this case SU(2) doublet fermions, may indeed be observable at the LHC, since they can be produced by Drell-Yan processes and lead to displaced vertices when they decay. However, the direct observation of lepton number violating processes is on the edge at LHC.

\vspace*{.5cm}

\end{titlepage}

\newpage

\renewcommand{\thefootnote}{\arabic{footnote}}
\setcounter{footnote}{0}


\section{Introduction}

From neutrino oscillations, it is evident that neutrinos are massive, see, \eg,  \Refs~\cite{GonzalezGarcia:2007ib,Schwetz:2011zk}. If the neutrino masses originate from physics beyond the Standard Model (SM) and are suppressed by a high energy scale, it is convenient to parameterize the impact of the heavy fields, present in the high-energy theory,  by a tower  of effective  operators $\mathcal{O}^{d}$ of dimension $d>4$. These operators made out of the SM fields, are invariant under the SM gauge group~\cite{Weinberg:1979sa,Wilczek:1979hc} (see also \Ref~\cite{Buchmuller:1985jz}).
The operator coefficients are weighted by inverse powers of the scale of new physics $\Lambda_{\mathrm{NP}}$: 
 \be\label{equ:leff}
\mathcal{L} = \mathcal{L}_{\rm SM} + \mathcal{L}^{d=5}_{\text{eff}} 
+ \mathcal{L}^{d=6}_{\text{eff}} + \cdots
\, , \quad \textrm{with} \quad \mathcal{L}^{d}_{\text{eff}} \propto \frac{1}{\Lambda_{\mathrm{NP}}^{d-4}} \, \mathcal{O}^{d}\,.
\ee
Some of these effective operators result in corrections to the
 low-energy SM parameters and in exotic couplings.
It is also known that there is only one possible operator at the lowest order in the expansion, $\mathcal{L}^{d=5}_{\trm{eff}}$,  namely, the famous Weinberg operator~\cite{Weinberg:1979sa},
\be
\mcl{O}_W = (\overline{L^{c}} {\rm i} \tau^{2} H)\, (H {\rm i} \tau^2 L)
\ee
which leads, after Electroweak Symmetry Breaking (EWSB), to Majorana masses  for the neutrinos. Here $L$ and $H$ stand for the SM
 lepton and Higgs doublets, respectively.  At tree level, $\mcl{O}_W$ can only be mediated in three ways \cite{Ma:1998dn}:
by a singlet fermion, a triplet scalar, or a triplet fermion, leading to the famous type~I~\cite{Minkowski:1977sc,Yanagida:1979as,GellMann:1980vs,Mohapatra:1979ia}, type~II~\cite{Magg:1980ut,Schechter:1980gr,Wetterich:1981bx,Lazarides:1980nt,Mohapatra:1980yp,Cheng:1980qt}, and type~III\cite{Foot:1988aq} see-saw mechanisms, respectively (see also \Ref~\cite{Abada:2007ux}). For recent discussions of SUSY versions
see for example \cite{Esteves:2009vg,Nath:2010zj,Esteves:2010ff} and references therein.
 Compared to the electroweak scale, the mass of the neutrinos in all three cases appears suppressed by a factor 
$v/\Lambda_{\mathrm{NP}}$, where $v$ is the Vacuum Expectation Value (VEV) of the Higgs boson. Substituting typical values, one obtains that the original see-saw mechanisms are pointing towards the Grand Unified Theory (GUT) scale.

More recently, however, scenarios in which $\Lambda_{\mathrm{NP}} \sim
\mathrm{TeV}$ have been drawing some attention, since they are potentially
testable at the LHC. In  these cases, additional suppression
mechanisms for the neutrino masses are required, and several possibilities open up: For example, the neutrino mass may be generated radiatively, where the additional suppression comes from loop
integrals, or the smallness of the neutrino mass may be protected by lepton number, such as in the inverse see-saw. 
In this study, we instead argue that the $d=5$ operator in \equ{leff} is forbidden, and neutrino masses originate from higher dimensional operators~\cite{Babu:1999me,Chen:2006hn,Gogoladze:2008wz,Giudice:2008uua,Babu:2009aq,Gu:2009hu,Bonnet:2009ej,Picek:2009is,Liao:2010rx,Liao:2010ku,Liao:2010ny,Kanemura:2010bq} (see also \Refs~\cite{Gouvea:2007xp,Xing:2009hx,Cogollo:2010jw,Brignole:2010nh,Liao:2010cc,Kumericki:2011hf} for related discussions). Note that there may be additional suppression mechanisms at work in this approach, such as loop suppression or a small lepton number violating parameter, see \Ref~\cite{Bonnet:2009ej} for an example.

There are several key ingredients to neutrino masses from higher dimensional operators \cite{Bonnet:2009ej}. Consider, for instance, 
the operator 
\begin{equation}
\mathcal{O}^7 =  (LLHH)(H^\dagger H) \, , \label{equ:d7}  
\end{equation}
where we have omitted spin, flavor, and gauge indices.  In this case, the $(H^\dagger H)$ component can be closed in a loop, which means that the $d=5$ operator is generated radiatively, and the $d=7$ operator may not be the leading contribution to neutrino mass (depending on the new physics scale). We do not consider such operators in this work, which means that~\cite{Bonnet:2009ej}
\begin{enumerate}
\item
 We have to forbid the lower dimensional operators by a U(1) or discrete symmetry.
\item
 We need new (scalar) fields to construct the higher dimensional operators, since $(H^\dagger H)$ is a singlet under any such symmetry.
\end{enumerate}
The simplest possibilities to enhance the field content of the SM are the addition of a
Higgs singlet~\cite{Chen:2006hn,Gogoladze:2008wz}
\begin{align}
 \mathcal{L}^{d=n+5}_{\text{eff}} =  \frac{1}{\Lambda_{\mathrm{NP}}^{d-4}} (LLHH) (S)^{n} \, , \quad n=1,2,3, \hdots 
\label{equ:s}
\end{align}
or the addition of a Higgs doublet, leading to the Two Higgs Doublet Model (THDM)~\cite{Gunion:1989we,Babu:1999me,Giudice:2008uua,Bonnet:2009ej}
\begin{align} 
\mathcal{L}^{d=2n+5}_{\text{eff}} =  \frac{1}{\Lambda_{\mathrm{NP}}^{d-4}}
	(LLH_{u} H_{u}) (H_{d} H_{u})^{n} \, , \quad n=1,2,3, \hdots \, .
\label{equ:hh}
\end{align}

As it has been demonstrated in \Ref~\cite{Bonnet:2009ej} in the framework of the THDM, the decomposition
of  \equ{hh} often leads to a linear or inverse see-saw structure~\cite{GonzalezGarcia:1988rw,Mohapatra:1986bd,Nandi:1985uh} if two extra fermion singlet fields ${N_R}$ and ${N'_L}$ are involved. 
The neutral fermion mass matrix then reads, in the basis 
$\begin{pmatrix}
\nu_L^{c}
&
N_R
&
{N'_L}^{c}
\end{pmatrix}$,
\begin{equation}
 M_\nu = 
\begin{pmatrix}
 0 
 & (Y_{\nu}^{\sf T}) v
 & \epsilon (Y'^{\sf T}_{\nu})
 \\
 (Y_{\nu}) v
 & \mu'
 & \Lambda_{\mathrm{NP}}
 \\
 \epsilon (Y'_{\nu})
 & \Lambda_{\mathrm{NP}}^{\sf T}
 & \mu'{}'
\end{pmatrix} \, .
\label{equ:iss}
\end{equation}
Here $\epsilon$, $\mu'$, and $\mu'{}'$ are typically introduced ad hoc
as small parameters, because they are protected by lepton number (for
the $\epsilon$-term, see also
\Refs~\cite{Abada:2007ux,Gavela:2009cd}). However, if the neutrino
mass is generated by a higher than $d=5$ effective operator, some terms
in the neutral fermion mass matrix in \equ{iss} can only originate from
non-renormalizable interactions, which means that extra fields are
needed and that the lepton number violating parameters are suppressed
by powers of $\Lambda_{\mathrm{NP}}$. A general discussion and two
specific examples leading to the $\mu$- and $\epsilon$-term in
\equ{iss} can be found in \Sec~3 of \Ref~\cite{Bonnet:2009ej}.

In principle, all neutrino mass models  can be
supersymmetrized. In practice, however, it turns out that supersymmetry
often requires additional particles for consistency, mainly due to the boundary condition
that the superpotential has to be holomorphic.
In this work, we apply \Ref~\cite{Bonnet:2009ej} to the framework of
SUSY, more specifically, we start with the minimal supersymmetric standard model
(MSSM) and the next to minimal one (NMSSM). 
A related discussion in the NMSSM framework can be found in 
 \Ref~\cite{Gogoladze:2008wz}.
In \Sec~\ref{sec:hdim}, we
systematically discuss higher dimensional effective operators
including two Higgs doublets and one scalar within SUSY.  Then in
\Sec~\ref{sec:inverse}, we focus on one specific decomposition leading to
a linear or inverse see-saw in the form of \equ{iss}. 
We will show that this requires an
extension of the particle content of the model, which is potentially
observable at the LHC. The main features will be outlined using a specific
model, where we also discuss constraints
due to existing data. In \Sec~\ref{sec:lhc}, we show that 
 lepton flavor mixing related to neutrino physics
is potentially testable at the LHC.  We also show that the cross sections for
some lepton number
violating processes can be significantly larger than naively expected.
However, it turns out that they are on the edge of observability at
the LHC.
Finally, in \Sec~\ref{sec:summary} we 
draw our conclusions.  In
\App~\ref{app:uvcompl} we comment on possible fundamental theories
leading to the $d=7$ operator discussed in this paper. In 
\App~\ref{app:approxdiag} we approximate some of the couplings
for the model considered in \Sec~\ref{sec:lhc}. 

\section{Neutrino masses from higher than d=5 operators in SUSY}
\label{sec:hdim}

\begin{table}[t!]
\begin{center}
\begin{small}
\begin{tabular}{ccll}
\hline \hline
& Op.\#
&Effective interaction
& Charge \\
\hline
$d=5$
& 1
& $LLH_{u} H_{u}$
& $2q_{L} + 2q_{H_{u}}$
\\
$d=7$
& 3
& $LLH_{u} H_{u} H_{d} H_{u}$
& $2 q_{L} + 3 q_{H_{u}} + q_{H_{d}} $
\\
$d=9$
& 7
&  $LLH_{u} H_{u} H_{d} H_{u} H_{d} H_{u}$
& $2q_{L} + 4 q_{H_{u}} + 2 q_{H_{d}} $
\\
\hline \hline
\end{tabular}
\end{small}
\mycaption{\label{tab:opOverviewMSSM} Effective operators generating neutrino mass in the MSSM up to $d=9$. The operator numbers have been chosen in consistency with \Tab~\ref{tab:opOverviewNMSSM}.}
\end{center}
\end{table}

\begin{table}[t!]
\begin{center}
\begin{small}
\begin{tabular}{ccllc}
\hline \hline
& Op.\#
&Effective interaction
& Charge & Same as \\
\hline
$d=5$
& 1
& $LLH_{u} H_{u}$
& $2q_{L} + 2q_{H_{u}}$
\\
\hline
$d=6$
& 2
&$LLH_{u} H_{u} S$ 
& $2q_{L} + q_{H_{u}} - q_{H_d}$
\\
\hline
$d=7$
& 3
& $LLH_{u} H_{u} H_{d} H_{u}$
& $2 q_{L} + 3 q_{H_{u}} + q_{H_{d}} $
\\
& 4
& $LLH_{u} H_{u} S S$
& $2 q_{L} - 2 q_{H_{d}}$
\\
\hline
$d=8$
& 5
& $LLH_{u} H_{u} H_{d} H_{u} S$ 
& $2q_{L} + 2q_{H_{u}}$
& \#1
\\
& 6
& $LLH_{u} H_{u} S S S$ 
& $2q_{L} + 2q_{H_{u}}$
& \#1
\\
\hline
$d=9$
& 7
&  $LLH_{u} H_{u} H_{d} H_{u} H_{d} H_{u}$
& $2q_{L} + 4 q_{H_{u}}+ 2 q_{H_{d}} $
\\
& 8
&  $LLH_{u} H_{u} H_{d} H_{u} S S$
& $2q_{L} + q_{H_{u}} - q_{H_{d}}$
& \#2
\\
& 9
&  $LLH_{u} H_{u} S S S S$
& $2q_{L} + q_{H_{u}} - q_{H_{d}}$
& \#2 
\\
\hline \hline
\end{tabular}
\end{small}
\mycaption{\label{tab:opOverviewNMSSM} Effective operators generating neutrino mass in the NMSSM up to $d=9$. Here $S$ is the NMSSM scalar, which means that its charge $q_S$ is fixed by the terms $\lambda \hat S \hat H_u \hat H_d$, \ie, $q_S  = -(q_{H_u}+q_{H_d})$, and $\kappa \hat S^3$, \ie, $3 q_S=0$, which have been used to derive the charge condition (\cf, \equ{nmssm}).}
\end{center}
\end{table}

We take the MSSM as a starting point for various extensions.
Note that its Higgs sector has the structure of a type~II--THDM with the
restriction that it is CP-invariant at leading order.  As
in \Ref~\cite{Bonnet:2009ej}, we require a discrete symmetry (in the sense of a
matter parity) to forbid the $d=5$ operator as leading contribution to
neutrino mass, where the simplest possibility is $\mathbb{Z}_3$ in the
case of SUSY.\footnote{Note, that the requirement of forbidding the $d=5$ operator
automatically implies conserved R-parity because a $\Delta L=1$ operator
immediately implies a contribution to the  $d=5$ operator~\cite{Hall:1983id}. 
For example, the sneutrino can get a VEV  if R-parity is broken.
Because of the neutrino-sneutrino-neutralino interaction, an additional
$d=5$ effective operator, which contributes to the neutrino mass, is
then possible~\cite{Hirsch:2000ef}. Therefore, we require R-parity
conservation, which also has strong constraints on the possible
decompositions of the effective operators.} In
\Tab~\ref{tab:opOverviewMSSM}, we list all possible higher dimensional
operators made from lepton doublets and the two Higgs fields up to
dimension nine. Note that compared to the THDM, the holomorphicity of the 
superpotential implies that  the only possible
effective $d=7$ operator in the MSSM is $L L H_u H_u H_d H_u$, and it also
limits the  number of possible decompositions further.\footnote{The 
holomorphicity of the superpotential implies that interactions
among scalars of the form $\phi_i \phi_j \phi^\dagger_k$ can 
only be introduced via F-terms (another possibility
to get non-holomorphic terms for the neutrino mass
operator are non-canonical terms in the K\"ahler potential,
as discussed in \cite{Casas:2002sn,Brignole:2010nh}).
Since there are no SUSY invariant
interactions with both, fermions and F fields, the only possible
effective $d=7$ operator in the MSSM is $L L H_u H_u H_d H_u$. }
 In \Tab~\ref{tab:opOverviewMSSM},
we show in addition to the possible operators the required charge 
combination such that the corresponding operator respects the
discrete symmetry.
However, in the MSSM, the combination $\mu \hat H_u \hat H_d$ appears in
the superpotential, which is breaking such the discrete symmetry explicitly. 
For this reason, we consider  models with extended particle content.

A possible extension beyond the MSSM is the NMSSM, where an 
additional Higgs singlet $S$ is introduced, see \eg\
\Ref~\cite{Ellwanger:2009dp,Maniatis:2009re} for reviews. 
This singlet couples to the usual Higgs doublets $H_u$ and $H_d$
and obtains
a non-zero vacuum expectation value. The NMSSM
superpotential is
\begin{align}
  W_\text{NMSSM} = W_\text{Yuk} + \lambda \hat S \hat H_u \hat H_d + \kappa \hat S^3\,,
\label{equ:nmssm}
\end{align}
where $W_\text{Yuk}$ denotes the superpotential for the Yukawa
couplings, \ie, the MSSM superpotential.  In this case, the charge of
$q_S$ is fixed by the second and third terms being uncharged. 
From the discrete symmetry point of view, one can easily see that
the last term in \equ{nmssm} is invariant under the $\mathbb{Z}_3$
for any charge assignment. 

 In
\Tab~\ref{tab:opOverviewNMSSM}, we list all possible higher dimensional
neutrino mass operators made from lepton doublets and the two extra Higgs fields up to
dimension nine for the NMSSM. In the column ``Charge'' we also show the
discrete symmetry charge using the fact that the terms in \equ{nmssm}
have to be uncharged. In the last column ``Same as'', we
indicate if the same condition as for a lower dimensional operator is
obtained, \ie, the lower dimensional operator cannot be avoided in
this case.  One can read off the table that operators \#2, \#3, \#4, and 
\#7 can be independently chosen as leading contribution
of neutrino mass, while the lower dimensional operators are forbidden.
 In general one can
show that the NMSSM operators of the type $LLH_uH_u(H_dH_u)^n
S^k$ (with $n \geq 1$, $k\geq1$ or $n=0$, $k \geq 3$) always imply
that other operators of lower dimension are allowed as well. This is due to the fact
that one finds field products of the
type $H_uH_dS$ or $S^3$, which have to be singlets under the discrete
symmetry, since they appear in the superpotential \equ{nmssm}.
This means that $d>7$ effective operators generating neutrino mass 
with singlet scalars will always come together with lower dimensional
operators. On the other hand, the effective operators with lepton and Higgs doublets only (such 
as \#3 and \#7) are {\em per se} interesting alternatives because one can choose even higher 
dimensional operators $d \ge 9$ as leading contribution.

Note that operators \#1, \#2, and \#4 have been studied in
\Ref~\cite{Gogoladze:2008wz}, whereas we focus on the $d=7$ operator
\#3 in the following. 
In this case, a possible charge assignment for
the $\mathbb{Z}_3$ symmetry is
\begin{align}
  q_{H_u}=0,\ q_{H_d}=1,\ q_L=1,\ (q_S = 2)\,.
 \end{align}
 While in this case, both the MSSM and NMSSM can be used as a
 framework, one has to be aware of the fact that the $\mu$-term of
 the MSSM, $\mu \hat H_u \hat H_d$, explicitly breaks the discrete
 symmetry. This problem is automatically circumvented in the NMSSM, as
 \equ{nmssm} respects the discrete $\mathbb{Z}_3$ symmetry and
 generates the $\mu$-term when $S$ takes a VEV. In SUSY,
 there are some differences compared to the THDM case in \Ref~\cite{Bonnet:2009ej}.
 For instance, the Lagrangian in \Ref~\cite{Bonnet:2009ej} was invariant under a new
U(1) symmetry in some cases, taking the role of the $\mathbb{Z}_3$ symmetry here,
  which potentially lead to unwanted Goldstone bosons; see discussion in \Sec~3.1 of
 \Ref~\cite{Bonnet:2009ej}. Even if the
Lagrangian was invariant under such a symmetry, it is
 obvious that \equ{nmssm} would break it explicitly, while it 
 respects $\mathbb{Z}_3$. 
In addition, note that a $d=5$ operator is inevitably generated by 
connecting the external $H_d$ and $H_u$ lines of a $d=7$ operator 
using a discrete symmetry breaking term $m^2 H_d \cdot H_u$ 
(see discussion in \Sec~3.1 of \Ref~\cite{Bonnet:2009ej}). 
The term $\mu \hat H_u \cdot \hat H_d$ in the superpotential corresponds to the scalar 
terms $|\mu|^2 H_u^\dagger H_u$ and $|\mu|^2 H_d^\dagger H_d$ in the Lagrangian here
(see \eg\ Sec. 16 of Ref. \cite{Aitchison:2005cf}), which means that
this problem does not occur. 
Instead, one 
 has a SUSY soft breaking term $B \mu  H_u \cdot H_d$, which however 
would break the discrete symmetry and thus occurs only below
 $\Lambda_{\mathrm{NP}}$. Therefore it should be
 sufficiently smaller than $\Lambda_{\mathrm{NP}}$ resulting in a
suppressed $d=5$ one-loop contribution. 
Note however, that its value is bounded
from below due to searches for the MSSM Higgs boson as it is proportional
to the mass of the pseudoscalar Higgs boson. 

The possible decompositions of $L L H_u H_u H_d H_u$ are
 systematically derived in \App~\ref{app:uvcompl} at tree level, where
 the mediators for different possibilities are listed in
 \Tab~\ref{tab:med}. Note that the right-handed fields listed there
have to be incorporated as charge-conjugated left-handed fields in
the superpotential. 
 These decompositions can be roughly categorized as extensions of
 the well known $d=5$ decompositions, the type I, II, and III see-saw
 scenarios. We define a decomposition as extended type II see-saw, if
 all mediators are scalars, \ie, decompositions \#5, \#6, and
 \#21-\#24 in \Tab~\ref{tab:med}. Therefore the only appearing
 fermions are the external lepton doublets. The only lepton number
 violating interaction is then
\begin{align}
	(\overline{L^c}\ii\tau_2\vec\tau L)\vec \phi\,,
\end{align}
where $\phi$ represents one of the scalar mediators. This vertex
violates lepton number by $\Delta L = 2$ and therefore conserves
R-Parity.  All other decompositions that have fermionic mediators can
be seen as extensions of the $d=5$ type I or type III see-saw
mechanism. Since we can have several combinations of scalar fields and
SU(2) singlet, doublet, or triplet fields as mediators, a further
distinction will not be made.  Depending on the topology and the
actual realization of these operators, the various decompositions have
different characteristics. Note that
integrating out all but two neutral fermion fields will lead to an
inverse see-saw-like scenario, as in \equ{iss}.

\begin{figure*}[tb]
\begin{center}
\includegraphics[width=.4\linewidth]{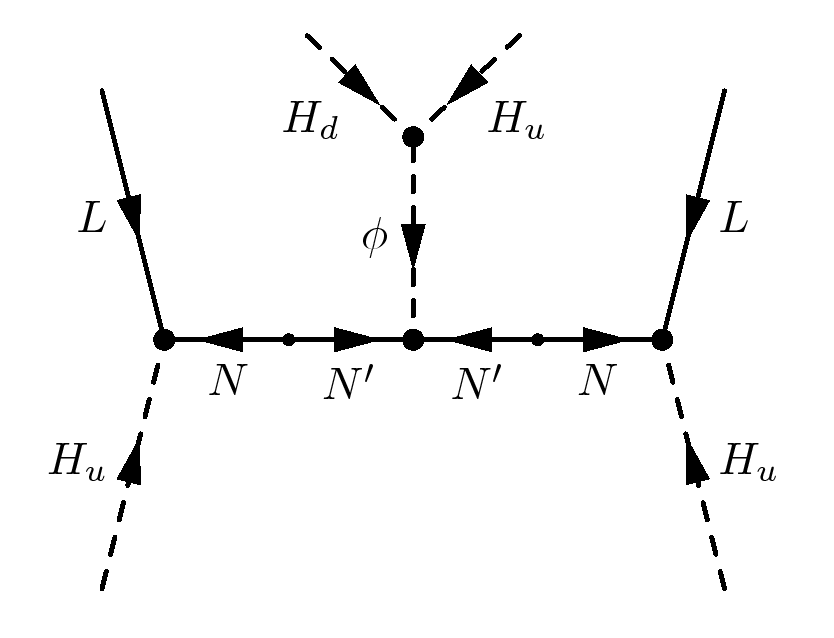}
\end{center}

\mycaption{\label{fig:d7op1}
Decomposition \#1 from \Tab~\ref{tab:med} of the effective $d=7$ operator $L L H_u H_u H_d H_u$. Here $N$ and $N'$ are fermion singlets, and $\phi$ is a scalar singlet.}
\end{figure*}

\begin{figure*}[t]
\begin{center}
\includegraphics[width=.4\linewidth]{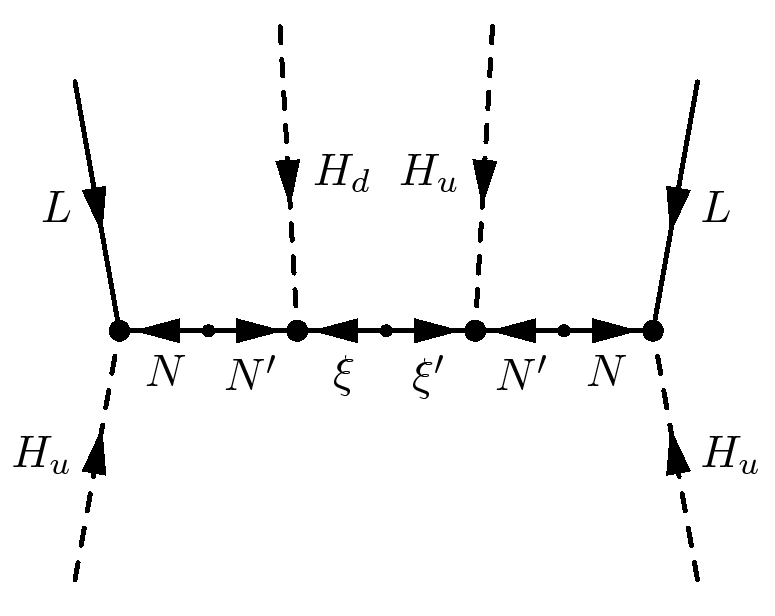}
\end{center}
\mycaption{\label{fig:d7op17} Decomposition \#17 from \Tab~\ref{tab:med} of the effective $d=7$ operator $L L H_u H_u H_d H_u$. Here $N$ and $N'$ are fermion singlets, and $\xi$ and $\xi'$ are fermion doublets.}
\end{figure*}

Let us now illustrate some of the complications involving extra scalars as mediators.
As an example we take decomposition \#1 from \Tab~\ref{tab:med}, shown in
\figu{d7op1}.  One can easily see that the scalar $\phi$ has the same quantum numbers as the
scalar of the NMSSM, and it also has the same coupling to the Higgs
fields. If the terms in \equ{nmssm} are present in the superpotential,
it can get a VEV $v_\phi$. This in turn means that the $d=6$ operator of the type
$LLH_uH_uS$ (\#2 in \Tab~\ref{tab:opOverviewNMSSM}) may be allowed, where $\phi \equiv S$,
which may be the leading contribution to neutrino mass. Indeed one can see 
from the $\lambda H_u H_d \phi^\dagger$-vertex in \figu{d7op1} that 
 we have for the discrete symmetry charge $q_{\phi}=q_{H_u}+q_{H_d}$, 
which means that we cannot forbid the $d=6$ operator $LLH_uH_uS$ 
which leads to neutrino mass if $\phi$ obtains a VEV.
In summary, the MSSM extended by a scalar singlet mediator can potentially
be NMSSM-like, and can potentially induce the $d=6$ operator, which may
dominate the neutrino mass contribution.
Since the operator $LLH_uH_uS$ in the NMSSM has been studied in
\Ref~\cite{Gogoladze:2008wz}, and since we want to avoid the $d=6$
operator genuinely, we focus on decompositions with two neutral
fermions (to reproduce the inverse see-saw) and no singlet scalars in
the following. One of the simplest examples is decomposition \#17 in
\Tab~\ref{tab:med}, which we will discuss in greater detail, see \figu{d7op17}. While
for neutral SM singlets as mediators the production rates of the new
particles are rather low, the SU(2) doublets in \#17 will lead to
gauge interactions with potentially observable phenomenology at the
LHC. However, note that also the fermion singlets could be replaced by
triplets, which would lead to a see-saw III-type phenomenology.

\section{A linear or inverse see-saw example}
\label{sec:inverse}

We have seen in the previous section that
extensions of the MSSM containing NMSSM-like singlets do have
some problems. Therefore we consider a model where we add two
gauge singlet superfields $\hat N$ and $\hat N'$
and a vector-like pair of $SU(2)$ doublets $\hat \xi$ and $\hat \xi'$
with hypercharges $Y = \frac{1}{2}$ and $Y = -\frac{1}{2}$, respectively.
 Note, that the holomorphicity
of the superpotential requires to use left-handed fields only and, thus,
$\hat \xi$ corresponds to the charge conjugated field {\bf 2}$^R_{-1/2}$
of decomposition \#17 in table \ref{tab:med}.
Moreover, none of these additional fields can participate in the
breaking of the electroweak symmetry as we require R-parity conservation. 
The corresponding superpotential  is
\begin{align} \label{equ:W17}
\begin{split}
 W = \ & W_\text{quarks} + Y_e \hat{e}^c \hat L \cdot \hat H_d
 - Y_N \hat N \hat L \cdot \hat H_u
 + \kappa_1 \hat N' \hat \xi \cdot \hat H_d
 - \kappa_2 \hat N' \hat \xi' \cdot \hat H_u
+ m_N \hat N \hat N' \\ &
+ m_\xi \hat \xi' \cdot \hat \xi
+ \mu \hat H_u \cdot \hat H_d\,,
\end{split}
\end{align}
and the corresponding lepton number assignments are
\begin{equation}
L(\hat N) = -1 \,,\,\, L(\hat N') = +1 \,,\,\,
 L(\hat \xi) = -1 \,,\,\, L(\hat \xi') = +1
\end{equation}
implying that the interaction proportional to $\kappa_2$ breaks
lepton number by two units\footnote{Note that this assignment
is, to some extent, arbitrary, and that a different assignment could be chosen
such that  the interaction proportional to $\kappa_1$ breaks
lepton number by two units. However, none of our conclusions
is affected by this specific choice.}.
This superpotential yields the following part of the Lagrangian for
the fermions carrying lepton number
\begin{align}
\begin{split}
 \mathcal{L}^\text{fermionic} =\ & 
 -Y_e (e^c L \cdot H_d + \widetilde{e}^*_R L \cdot \widetilde H_d 
 + e^c \widetilde L \cdot \widetilde H_d)
 + Y_N (N L \cdot H_u + \widetilde N L \cdot \widetilde H_u + N \widetilde L \widetilde H_u)\\
 &- \kappa_1 (N' \xi \cdot H_d + \widetilde N' \xi \cdot \widetilde H_d + N' \widetilde \xi \cdot \widetilde H_d )
 + \kappa_2 (N' \xi' \cdot H_u + \widetilde N' \xi' \cdot \widetilde H_u + N' \widetilde \xi' \cdot \widetilde H_u )\\
 &- m_N N' N
 - m_\xi \xi' \cdot \xi +\hc \, ,
 \end{split}
\end{align}
 using 2-component spinors. For completeness, note that for the leptons we use the ``$\sim$'' for
scalars, whereas for the Higgs bosons we use  the ``$\sim$'' for the fermionic partners.
After the Higgs fields get a VEV, the 
 mass matrix for the neutral fermions  reads in the basis
\begin{align}
f^0 = (\nu, N, N',\xi^0, {\xi'}^0)
\end{align}
\begin{align}
\begin{split}
 &M^0_f =
\left(\begin{array}{ccccc}
	0			& Y_N v_u	        & 0		        & 0			& 0	\\
	Y_N^{\sf T} v_u		& 0			& m_N^{\sf T}	        & 0			& 0	\\
	0			& m_N		        & 0		        & \kappa_1 v_d	        & \kappa_2 v_u	\\
	0			& 0			& \kappa_1^{\sf T} v_d	& 0			& -m_\xi	\\
	0			& 0			& \kappa_2^{\sf T} v_u	& -m_\xi		        & 0			
\end{array}\right)\,.
\end{split}
\label{equ:mass0}
\end{align}
The corresponding mass eigenstates will be denoted by $n_i$ 
 with $|m_i| \le |m_j|$ for $i< j$.
The mass terms for the charged fermions  are given by
\begin{align}
- v_d e^c Y_e e_L - m_\xi \xi^+ \xi'^-\,.
\end{align}
We can now determine the neutrino mass, by integrating out the mediator fields. For the sake of simplicity, let us first of all ignore the flavor structure. Integrating out the heavy doublets, we obtain
\begin{align}
 \mathcal L_N = N^\dagger \ii \bar\sigma^\mu \partial_\mu N + {N'}^\dagger \ii \bar\sigma^\mu \partial_\mu N' - m_N N N' - Y_N N L \cdot H_u - \frac{\kappa_1 \kappa_2}{m_\xi} N' N' H_u \cdot H_d + \hc \, ,
\end{align}
which reads in the basis $(\nu, N, N')$ after electroweak symmetry breaking
\begin{align}
{M^0_f}' = \begin{pmatrix}
0	& Y_N \, v_u	& 0 \\
Y_N \, v_u	& 0		& m_N \\
0	& m_N	& \hat \mu \\
\end{pmatrix}
\label{equ:issour}
\end{align}
with $\hat \mu = v_u v_d \, (2 \kappa_1 \kappa_2)/m_\xi $. This is an inverse see-saw mass matrix, as in \equ{iss}. As it is characteristic for the inverse see-saws from higher dimensional operators, the lepton number violating term is suppressed by a heavy scale. 
If in addition the singlet fermions are integrated out, we obtain for the neutrino mass scale
\begin{align}
 m_\nu = v_u^3 v_d Y_N^2 \frac{\kappa_1 \kappa_2}{m_\xi m_N^2} \, .
\label{equ:mnu}
\end{align}
For a neutrino mass $m_\nu \approx 1 \, \mathrm{eV}$ and $v \approx 177 \, \mathrm{GeV}$ and the heavy mass scale at $1 \, \mathrm{TeV}$ this means couplings $\mathcal O(10^{-3})$ are required. Note that this coupling strength in not unreasonably small, although these couplings are, in addition, protected by lepton number.  

If we assume instead a hierarchy of the heavy particles where the isospin singlets are heavier than the doublets, we first can integrate out the singlets. The mass matrix for the remaining neutral fields $(\nu, \xi^0, \xi'^0)$ reads then
\begin{align}
{M^0_f}'' = \begin{pmatrix}
0					& \tilde \kappa_1\, v_d	& \tilde \kappa_2\, v_u \\
\tilde \kappa_1\, v_d	& 0					& -m_\xi \\
\tilde \kappa_2\, v_u	& -m_\xi				& 0
\end{pmatrix}\,,
\end{align}
where $\tilde\kappa_{1/2}= \kappa_{1/2}\, Y_N^2 / m_N$. Integrating out the $\xi$ fields afterwards we again arrive at \equ{mnu} for the mass of the light neutrinos.

As in the conventional inverse see-saw, in which the $\hat\mu$-term in
\equ{issour} is introduced {\em ad hoc}, there are various interesting
phenomenological effects of this scenario. It is expected that
non-unitarity and its CP violation can be tested at possible long-baseline
neutrino oscillation experiments (see, \eg,
\Refs~\cite{Antusch:2009pm,Meloni:2009cg}). Furthermore, one may observe
lepton-flavor-violating (LFV) processes such as $\mu \rightarrow e
\gamma$. Lepton-number-violation, on the other hand, is expected to be
hardly testable in conventional scenarios, since the heavy Majorana
neutrinos form pseudo-Dirac particles with suppressed Majorana
character, see, \eg, \Ref~\cite{Ibarra:2010xw}\footnote{%
Some attempts to avoid the suppression are discussed in \eg, Refs.~\cite{Keung:1983uu,Kersten:2007vk,Blaksley:2011ey}.
}.

There are several ways to realize a flavor structure that is in
accordance with neutrino physics. Since there are three distinct
(active) mass eigenstates, at least two of them must be massive. The
straightforward approach is to add three generations of the heavy
fields, which leaves, however, many unconstrained parameters. Another
possibility
is to generate one neutrino mass by the inverse seesaw with one
generation of mediators, and the second neutrino mass at the one-loop
level~\cite{Hirsch:2009ra} if the flavor structures in the
soft SUSY sector differs from the ones in the superpotential. 
  A third version is the
minimal inverse seesaw scenario  in \Ref~\cite{Malinsky:2009df},
consisting of only two generations of the heavy fields, which narrows
down the number of free parameters.

We follow a similar approach, where we assume that one neutrino state is massless. We assume two generations of $N$ and $N'$ each, and only one generation for the other mediators. Thus, compared to \equ{mnu}, we obtain a mass matrix
\begin{align}
 	(m_\nu)_{\alpha\beta} = v_u^3 v_d (Y_N)_{\alpha i} (m_N^{-1})_{ij} \mu_{jk} (m_N^{-1,\T})_{kl} (Y_N^\T)_{l\beta}\,,
\end{align}
where
\begin{align}
 	\mu_{jk} = \frac{1}{m_\xi} \left((\kappa_1)_{j} (\kappa_2)_{k} + (\kappa_2)_{j} (\kappa_1)_{k}\right)\,.
\end{align}
The flavor basis can be chosen in a way that $M_N$ (and consequently $M_N^{-1}$) is diagonal, without loss of generality. We choose the parameters to reproduce tri-bimaximal mixings~\cite{Harrison:2002er}\footnote{If $\theta_{13}>0$, as indicated by the recent T2K hint~\cite{Abe:2011sj}, a different flavor structure can be easily implemented.}, which depend on the mass hierarchy:

{\bf Normal hierarchy.} 
A rather straightforward parameterization is:
\begin{align} \label{equ:fstruct_NH}
 	Y_N = y_N\begin{pmatrix}
 	       	\frac{1}{\sqrt{3}}	& 0 \\
 	       	\frac{1}{\sqrt{3}}	& -\frac{1}{\sqrt{2}} \\
 	       	\frac{1}{\sqrt{3}}	& \frac{1}{\sqrt{2}}
 	      \end{pmatrix}\,,\quad
 	\kappa_1 = k_1 \vectwo{-1}{1}\,,\quad
 	\kappa_2 = k_2 \vectwo{1}{1}\,,\quad
 	M_N = m_N \matrixtwo{1}{0}{0}{\rho}\,,
\end{align}
where 
\begin{align}
\rho = \sqrt{m_2/m_3} \, , \qquad 2 v_u^3 v_d y_N^2 k_1 k_2/(m_N^2 m_\xi) \stackrel{!}{=} m_2\,.
\label{equ:rho}
\end{align}
This reproduces the tri-bimaximal mixing pattern and two non-zero mass eigenvalues. In this case, the flavor structure of the neutrinos is dominantly  generated by the neutrino Yukawa couplings $Y_N$.  Since we have more parameters than constraints from neutrino physics, there is a certain freedom in the parameters of the couplings. For example one can vary $y_N$ as long as this is compensated by an according change of $m_N$ or $k_{1/2}$. The mass ratio $\rho = \sqrt{m_2/m_3}$ can be generated by $y_N$, $m_N$ or $k_{1/2}$. A possible set of parameters is $3y_N=10^{-3}$, $k_1=k_2=10^{-2}$, $\tan \beta = 10$, $m_N=1070 \, \mathrm{GeV}$, and $m_\xi=200 \, \mathrm{GeV}$, which we will use in the next section.
We have checked that this point
is compatible with bounds on  rare lepton decays such as $\mu\to e \gamma$
as well as with the search for the tri-lepton signal of supersymmetric
particles at the Tevatron \cite{Abazov:2009zi} and searches for new
physics in final states containing leptons at the LHC
 \cite{Chatrchyan:2011ff,Collaboration:2011dc}. In order to satisfy the bounds from the rare decays,
we have assumed that the scalar leptons are so heavy that their contributions are suppressed and that the leading contributions are due to loops containing fermions and the $W$-boson.
 Note that the product $v_u^3 v_d$ in \equ{rho} peaks at about $\tan \beta \simeq 2$, and it becomes small for large $\tan \beta$. 

{\bf Inverted hierarchy.} The inverted hierarchy can be obtained by the parameterization
\begin{align}
 	Y_N = y_N\begin{pmatrix}
 	       	\sqrt{\frac{2}{3}}	& \frac{1}{\sqrt{3}} \\
 	       	-\frac{1}{\sqrt{6}}	& \frac{1}{\sqrt{3}} \\
 	       	-\frac{1}{\sqrt{6}}	& \frac{1}{\sqrt{3}}
 	      \end{pmatrix}\,,\quad
 	\kappa_1 = k_1 \vectwo{-1}{1}\,,\quad
 	\kappa_2 = k_2 \vectwo{1}{1}\,,\quad
 	M_N = m_N \matrixtwo{1}{0}{0}{\rho}\,,
\end{align}
where $\rho = \sqrt{m_1/m_2}$ and 
\begin{align}
 2 v_u^3 v_d y_N^2 k_1 k_2/(m_N^2 m_\xi) \stackrel{!}{=} m_1\,.
\end{align}
In the following, we only consider the normal hierarchical example.

\section{LHC phenomenology}
\label{sec:lhc}

In many supersymmetric versions of neutrino mass models one finds
traces of the underlying mechanism generating neutrino masses in the
spectrum and decay properties of the supersymmetric particles, for
an incomplete list see e.g.\
\cite{Hisano:1998fj,Deppisch:2002vz,Deppisch:2004fa,Arganda:2005ji,%
Freitas:2005et,Buckley:2006nv,%
Deppisch:2007xu,Esteves:2009vg,DeCampos:2010yu,Esteves:2010ff,%
Esteves:2010si,Abada:2011mg,DeRomeri:2011ie}. 
Before discussing this in more detail, let us have a look on 
the number of parameters  related to neutrino physics in
this model. Working a basis where the lepton Yukawa couplings
are flavor diagonal, the following parameters contribute:
$Y_N$, $\kappa_i$, $m_N$ and $m_\xi$ which amounts 
in our specific model into 24 real parameters if all CP violating
phases are taken into account, but which gets reduced to 13 if
all phases are zero. From these at most six can be determined
in the near future, leaving 18 (7) parameters undetermined.
Here the question arises to which extent they might be measured
or at least constrained at the LHC. In principle we have sufficient
many decays to determine them: six heavy neutral states  decaying
into the three charged leptons, in total 18 decays.

\begin{figure}[t]
\begin{center}
 \includegraphics[width=.6\linewidth]{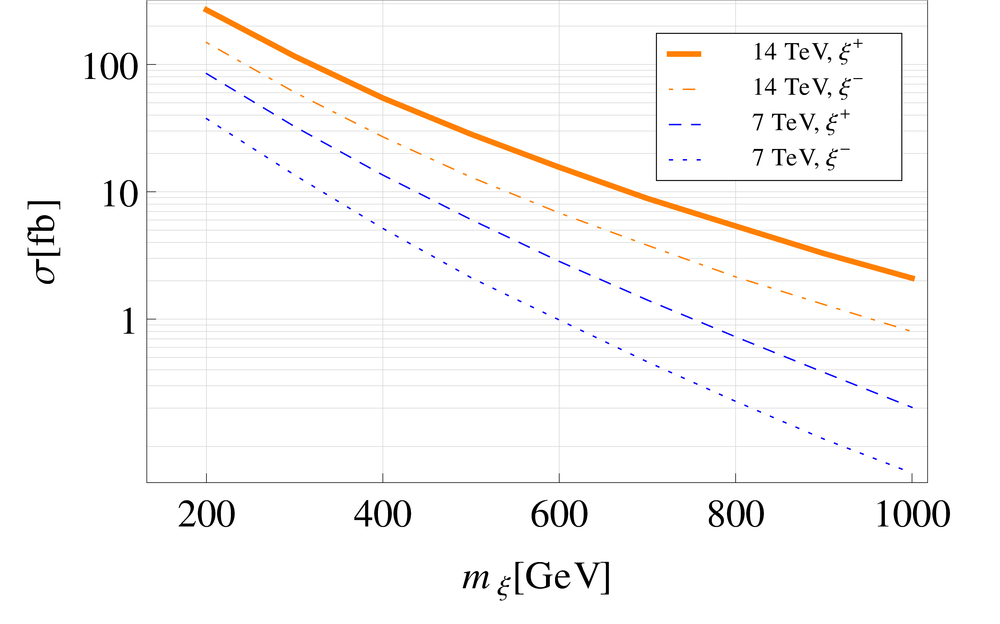}
\end{center}

\mycaption{\label{fig:Xsect_mxi} Total cross section $\sigma(pp\to \xi^\pm
\xi^0)$ as a function of the mass $m_\xi$.}
\end{figure}

Of particular interest is the question to which extent the new
particles, which are postulated in \equ{W17}, can be
produced at the LHC. It turns out that except for finely tuned
parameter combinations, the heavy states are either mainly $SU(2)$ singlets
or mainly $SU(2)$ doublets corresponding to the electroweak states. 
The singlet states contained in the superfields
$\hat N$ and $\hat N'$ can only be produced in cascade decays. However
this will happen in rare occasions only due to the smallness of the
involved Yukawa couplings. The $SU(2)_L$ doublets contained in
$\hat\xi$ and $\hat\xi'$, on the other hand, can in principle be
produced directly in Drell-Yan processes similarly to sleptons or
charginos and neutralinos within the usual MSSM \cite{Eichten:1984eu}.
In Figure~\ref{fig:Xsect_mxi} we show the cross section
for $\xi^+ \xi^0$ as a function of $m_\xi$ for $\sqrt{s}=$ 7~TeV and 14~TeV
assuming that the mixing with the singlet fields is small.
The numbers have been obtained using the {\tt WHIZARD} package
\cite{Kilian:2007gr}. 
If one takes for example a mass of 200~GeV for $\xi^+$ and $\xi^0$
using the numerical values for the couplings as given in the previous
section we find for the total cross section $\sigma(pp\to \xi^\pm
\xi^0)$ about 122 fb (417 fb) in case of 7 (14) TeV cms-energy
using the {\tt WHIZARD} program \cite{Kilian:2007gr}. Note, that 
the states $n_4$ and $n_5$ are mainly a nearly maximal
mixed superposition of the neutral components of the $SU(2)$ doublets
 in this case $n_{4,5} \simeq (\xi^0 \pm  \xi^0{}')/\sqrt{2}$.

In the following, we concentrate on the fermionic states, as they are
directly related to neutrino physics. We assume for the moment 
that their scalar partners are much heavier, so that the only possible decay
channels are into SM fermions and Higgs bosons, or vector bosons. In this
case, the main decay modes are
\begin{equation}
\xi^+ \to W^+ \nu_k \,,\,\, H^+ \nu_k
\end{equation}
for $\xi^+$, which is the Dirac fermion composed of the charged components
of $\xi$ and $\xi'$. 
Note, that there are no decays into $Z$ or $h^0$ as this
particle does not mix with the charged leptons at tree-level. 
One expects small
decay widths as these decays have their origin in the mixing of
$\xi^0$ with neutrinos and, thus, their widths are proportional to the
corresponding mixing matrix element squared. We indeed find 
$\Gamma(\xi^+) = 1.42 \ex{-5}$ keV.
In case of the neutral
fermions $n_i$, a larger variety of decay channels is possible:
\begin{eqnarray}
n_i &\to & W^\pm l^\mp_j \,,\,\, H^\pm l^\mp_j \\
n_i &\to & Z \nu_k  \,,\,\,h^0 \nu_k  \,,\,\,H^0 \nu_k  \,,\,\,A^0 \nu_k 
\end{eqnarray}
with $l_j=e,\mu,\tau$.  In these cases, the decays also originate from
the mixing of the neutral states with the neutrinos and, thus, the
corresponding widths are expected to be small as can be seen
in \Tab~\ref{tab_dw_allN} where we give the corresponding widths and branching ratios for the scenario discussed in the previous section.
  Obviously
some of the widths are so small that one can expect sizable decay
lengths at the LHC in the range of 100 $\mu$m to several mm once the
boost factor is taken into account. This is an important feature
because in this way one can not only suppress the SM background, but one
can also identify the leptons coming from these decays and
distinguish them from the ones coming from the cascade decays
of supersymmetric particles. Another interesting feature is, that
there are two pairs of states where within each pair the branching ratios
are nearly equal: $n_6/n_7$ and $n_8/n_9$. The reason is that they
form a quasi Dirac fermion. Also in case of $n_4$ and $n_5$ we have
a quasi Dirac fermion resulting in difficulties to determine
the branching ratios of the individual states. As a consequence
at most 9 branching ratios can be related to neutrino physics in praxis.

The fact that the $n_i$ decay into $W^\pm l^\mp$ clearly proves
that these particles carry lepton number and, thus, one might suspect
that they are related to the generation of neutrino masses. An
important question is in this context to which extent one can prove
their Majorana nature by observing both lepton charges in the final
states. Therefore one has to look for final states which violate lepton
number by two units compared to the initial one:
\begin{eqnarray}
u \bar{d} &\to &  l^+ l'{}^+ W^-\label{equ:Wll} \\
u \bar{d} &\to &   l^+ l'{}^+ W^- Z\label{equ:ZWll}\\
q \bar{q} &\to &  l^+ l'{}^+ W^-  W^-,  l^- l'{}^- W^+  W^+ \label{equ:WWll}
\end{eqnarray}
Note that the leptons can easily be of different generations due to the large mixing
angles in the neutrino sector.
In the calculation of these processes we have included all possible
intermediate particles to account for possible mixing effects, 
\eg, due to the pseudo-Dirac nature of the heavy additional neutral
fermions, which turn out to be important in case of  lepton number violating
final states. The interesting part of the Lagrangian is given by
\begin{equation}
\mathcal{L}_{Wl_jn_i} = 
-\frac{g}{\sqrt{2}}\bar{n}_i \gamma^\mu (a_{ij} P_L + b_{ij} P_R) l_j W^+_\mu 
-\frac{g}{\sqrt{2}}\bar{l}_j \gamma^\mu (a^*_{ij} P_L + b^*_{ij} P_R) n_i W^-_\mu 
\label{equ:lag_nlw}
\end{equation}
\begin{equation}
\mathcal{L}_{W\xi^+n_i} = 
-\frac{g}{\sqrt{2}}\bar{n}_i \gamma^\mu (c_{i} P_L + d_{i} P_R) \xi^- W^+_\mu 
-\frac{g}{\sqrt{2}}\bar{\xi}^- \gamma^\mu (c^*_{i} P_L + d^*_{i} P_R) n_j W^-_\mu 
\label{equ:lag_nxiw}
\end{equation}
where 
\begin{align} \label{eq_cpl1}
 a_{ij} = U_{ij}\,, \qquad b_{ij} = 0 \qquad j = e,\mu,\tau\,,
\end{align}
since these couplings have their origin in the left-handed couplings of
the SM leptons to the $W$-boson. Here we are working in a basis
where the Yukawa matrix of the charged leptons is diagonal.
Note, that $b_{ij}$ would only be
non-zero if there were a mixing between $\xi^-$ with the charged leptons.
The couplings to the $\xi^- = (\xi^-_L, (\xi^+)^c_R)^\T$ originate 
from the $SU(2)$ couplings between the $\xi$ doublet components
and are given by
\begin{align} \label{eq_cpl2}
 c_i = U_{i\xi'} \,, \qquad d_i = -U_{i\xi}^*\,.
\end{align}
Here $U$ denotes the matrix diagonalizing the mass matrix of the neutral
fermions, see \equ{mass0}.

\begin{table}[t]
\begin{center}
	\begin{tabular}{l|c|ccccc}
	\hline
	\rule{0pt}{1.1em}
	 Particle	& $\Gamma$[keV] & BR($W^\pm e^\mp$) & BR($W^\pm \mu^\mp$) & BR($W^\pm \tau^\mp$) &  BR($Z \nu$) &   BR($h^0 \nu$)\\
	 \hline \hline
	 \rule{0pt}{1.1em}
	 $n_4$	& $2.3\ex{-5}$ & $6.6\ex{-3}$ & $7.0\ex{-2}$ & $0.18$ & $0.36$ & $0.38$ \\
	 \hline
	 \rule{0pt}{1.1em}
	 $n_5$	& $1.9\ex{-5}$ & $1.2\ex{-2}$ & $0.41\ex{-2}$ & $0.18$ & $0.42$ & $0.34$\\
	 \hline
	 \rule{0pt}{1.1em}
	 $n_6$	& $1.2$  & $1.2\ex{-11}$ & $0.21$ & $0.21$ & $0.21$ & $0.37$\\
	 \hline
	 \rule{0pt}{1.1em}
	 $n_7$	& $1.2$ & $1.2\ex{-11}$ & $0.21$ & $0.21$ & $0.21$ & $0.37$ \\
	 \hline
	 \rule{0pt}{1.1em}
	 $n_8$	& $2.9$ & $0.14$ & $0.14$ & $0.14$ & $0.20$ & $0.38$  \\
	 \hline
	 \rule{0pt}{1.1em}
	 $n_9$	& $2.9$ & $0.14$ & $0.14$ & $0.14$ & $0.20$ & $0.38$ \\	
	 \hline \hline	 
	\end{tabular}
\mycaption{Total decay widths of the neutral mass eigenstates and branching ratios into the possible final states (where $\nu$ is the sum over the three light neutrino mass eigenstates).}
\label{tab_dw_allN}
\end{center}
\end{table}

\begin{figure}
 \begin{center}
\includegraphics[width=.4\linewidth]{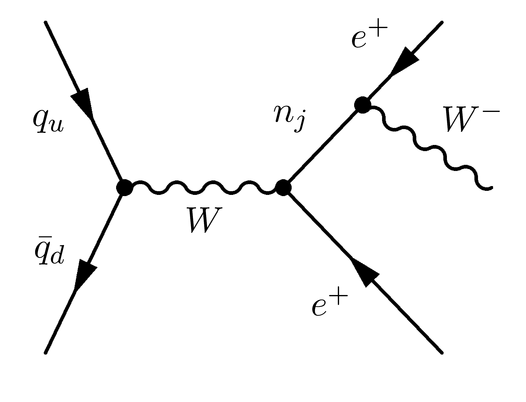}
\hspace{0.15\linewidth}
\includegraphics[width=.4\linewidth]{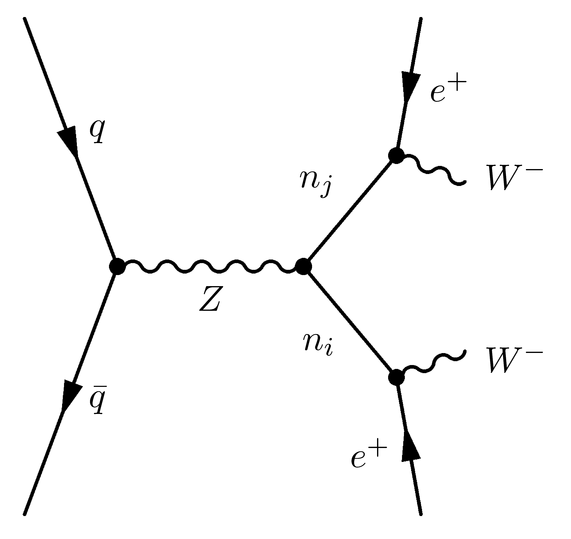}
\end{center}
\mycaption{\label{fig:feyn_p1} Dominant contribution to the
LNV processes  $u \bar{d} \to W^- e^+ e^+$ and 
$q \bar{q} \to W^- W^- e^+ e^+$.}
\end{figure}

The results for the $2 \to 3$  processes are shown in \Tab~\ref{tab_results1}. Note that in this case
only final states containing a $W$-boson are possible, as the
$\xi^+$ does not decay into charged leptons.
The main contributions in this case
are due to
\begin{equation}
u \bar{d} \to  l^+ n_i^* \,\,(i=1,\dots 5)
\end{equation}
as shown in Fig.~\ref{fig:feyn_p1}.
Here the $n_i$ are in this case either mainly neutrinos ($i=1,2,3$),
or an admixture of $\xi^0$ or
$\xi^0{}'$ ($i=4,5$).
The contributions of the neutrino-like states are suppressed because
they are off-shell, whereas in case of the $\xi^0/\xi^0{}'$-like states,
there are on-shell contributions. These are, however, somewhat
suppressed by the
small mixing elements with the neutrinos. We have put a cut on the
invariant mass of the leptons of 10 GeV as otherwise the $e^+ e^-$
final states would be enhanced by several orders of magnitude due to
an nearly on-shell photon.  The flavor mixed
final states are of the order of a few $fb$ and thus are potentially
observable if sufficient luminosity is accumulated. Note that 
for extracting the corresponding signal, only the hadronic final states
of the $W$-boson should be considered and, thus, the cross section
shown has to be multiplied by the corresponding branching ratio
$BR(W\to q \bar{q}')$. In the case that the two leptons have different flavor,
these processes are essentially background free. However, in case
of equal flavor leptons multi $W$-production in association with
a $Z$-boson or an off-shell photon will contribute. Both contributions
can be suppressed by putting cuts on the invariant mass of the two
leptons.

One also sees
from these tables that the lepton number violating final states are
strongly suppressed which is due to the appearance of the pseudo-Dirac
like state $n_4/n_5$ implying that the final contribution to the
cross section is proportional to $m^2_{n_5}-m^2_{n_4} \simeq O(m^2_\nu)$,
and thus tiny.

\begin{table}[tp]
 \begin{center}
   \begin{tabular}{c|ccr}
   \hline
    \rule{0pt}{1.2em}
 Process & $\sigma$ [fb] (7 TeV) & $\sigma$ [fb] (14 TeV)   \\
 \hline \hline
 \rule{0pt}{1.2em}
 $p p \rightarrow W^+ e^+ e^-$ 	&	$(1.651\pm0.024)\ex{2}$ 	&	$(4.161\pm0.023)\ex{2}$\\
   \rule{0pt}{1.2em}
 $p p \rightarrow W^- e^+ e^-$ 	&	$(9.240\pm0.033)\ex{}$		&	$(2.671\pm0.042)\ex{2}$\\
 \rule{0pt}{1.2em}
 $p p \rightarrow W^+ e^+ \mu^-$&	$(1.068\pm0.099)$		&	$(2.848\pm0.011)$\\
\rule{0pt}{1.2em}
 $p p \rightarrow W^+ e^- \mu^+$&	$(1.057\pm0.013)$		&	$(2.871\pm0.012)$\\
 \rule{0pt}{1.2em}
 $p p \rightarrow W^- e^+ \mu^-$&	$(5.748\pm0.015)\ex{-1}$	&	$(1.742\pm0.015)$\\
 \rule{0pt}{1.2em}
$p p \rightarrow W^- e^- \mu^+$&	$(5.755\pm0.015)\ex{-1}$	&	$(1.753\pm0.017)$\\
 \rule{0pt}{1.2em}
 $p p \rightarrow W^+ e^+\tau^-$&	$(1.058\pm0.096)$		&	$(2.861\pm0.011)$\\
\rule{0pt}{1.2em}
 $p p \rightarrow W^+ e^-\tau^+$&	$(1.056\pm0.095)$		&	$(2.854\pm0.011)$\\
 \rule{0pt}{1.2em}
 $p p \rightarrow W^- e^+\tau^-$&	$(5.714\pm0.015)\ex{-1}$	&	$(1.754\pm0.015)$\\
 \rule{0pt}{1.2em}
 $p p \rightarrow W^- e^-\tau^+$&	$(5.750\pm0.015)\ex{-1}$	&	$(1.744\pm0.019)$\\
 \rule{0pt}{1.2em}
 $p p \rightarrow W^+\mu^+\mu^-$&	$(1.676\pm0.014)\ex{2}$		&	$(4.116\pm0.023)\ex{2}$\\
 \rule{0pt}{1.2em}
 $p p \rightarrow W^-\mu^+\mu^-$&	$(9.242\pm0.033)\ex{}$		&	$(2.677\pm0.035)\ex{2}$\\
 \rule{0pt}{1.2em}
 $p p \rightarrow W^+\mu^+\tau^-$&	$(2.668\pm0.024)\ex{-1}$	&	$(7.092\pm0.028)\ex{-1}$\\
\rule{0pt}{1.2em}
 $p p \rightarrow W^+\mu^-\tau^+$&	$(2.652\pm0.026)\ex{-1}$	&	$(7.187\pm0.029)\ex{-1}$\\
 \rule{0pt}{1.2em}
 $p p \rightarrow W^-\mu^+\tau^-$&	$(1.432\pm0.006)\ex{-1}$	&	$(4.424\pm0.038)\ex{-1}$\\
 \rule{0pt}{1.2em}
 $p p \rightarrow W^-\mu^-\tau^+$&	$(1.439\pm0.004)\ex{-1}$	&	$(4.433\pm0.037)\ex{-1}$\\
 \rule{0pt}{1.2em}
 $p p \rightarrow W^+\tau^+\tau^-$&	$(1.665\pm0.023)\ex{2}$		&	$(4.138\pm0.063)\ex{2}$\\
 \rule{0pt}{1.2em}
 $p p \rightarrow W^-\tau^+\tau^-$&	$(9.265\pm0.034)\ex{}$		&	$(2.652\pm0.035)\ex{2}$\\
 \hline
 \rule{0pt}{1.2em}
 $p p \rightarrow W^- e^+ e^+$ 	&	$(4.711\pm0.069)\ex{-12}$ 	&	$(4.847\pm0.030)\ex{-11}$\\
 \rule{0pt}{1.2em}
 $p p \rightarrow W^+ e^- e^-$ 	&	$(1.423\pm0.008)\ex{-12}$	&	$(1.818\pm0.071)\ex{-11}$\\
 \rule{0pt}{1.2em}
 $p p \rightarrow W^- e^+ \mu^+$&	$(1.017\pm0.014)\ex{-11}$	&	$(9.869\pm0.054)\ex{-11}$\\
\rule{0pt}{1.2em}
 $p p \rightarrow W^+ e^- \mu^-$&	$(3.184\pm0.015)\ex{-12}$	&	$(3.22\pm0.15)\ex{-11}$\\
\rule{0pt}{1.2em}
 $p p \rightarrow W^- e^+\tau^+$&	$(1.169\pm0.015)\ex{-11}$	&	$(1.050\pm0.054)\ex{-10}$\\
\rule{0pt}{1.2em}
 $p p \rightarrow W^+ e^-\tau^-$&	$(4.173\pm0.020)\ex{-12}$	&	$(4.12\pm0.28)\ex{-11}$\\
\rule{0pt}{1.2em}
 $p p \rightarrow W^-\mu^+\mu^+$&	$(5.861\pm0.082)\ex{-9}$	&	$(2.278\pm0.013)\ex{-8}$\\
\rule{0pt}{1.2em}
 $p p \rightarrow W^+\mu^-\mu^-$&	$(2.377\pm0.010)\ex{-9}$	&	$(1.153\pm0.017)\ex{-8}$\\
\rule{0pt}{1.2em} 
 $p p \rightarrow W^-\mu^+\tau^+$&	$(1.184\pm0.013)\ex{-8}$	&	$(4.584\pm0.023)\ex{-8}$\\
\rule{0pt}{1.2em}
$p p \rightarrow W^+\mu^-\tau^-$&	$(4.788\pm0.018)\ex{-9}$	&	$(2.363\pm0.039)\ex{-8}$\\
\rule{0pt}{1.2em}
 $p p \rightarrow W^-\tau^+\tau^+$&	$(5.956\pm0.080)\ex{-9}$	&	$(2.292\pm0.031)\ex{-8}$\\
 \rule{0pt}{1.2em}
 $p p \rightarrow W^+\tau^-\tau^-$&	$(2.383\pm0.010)\ex{-9}$	&	$(1.120\pm0.014)\ex{-8}$\\
 \hline \hline
 \end{tabular}
\mycaption{Cross-sections for the processes with $W^\pm \ell^\pm \ell^\pm$ as final states (lepton number violating processes in lower section). A cut on the invariant lepton mass of 10 GeV has been assumed.}
\label{tab_results1}
\end{center}
\end{table}

\begin{table}[t]
 \begin{center}
   \begin{tabular}{c|cc}
   \hline
    \rule{0pt}{1.2em}
 Process & $\sigma$ [fb] (7 TeV) &  $\sigma$ [fb] (14 TeV)  \\
 \hline \hline
\rule{0pt}{1.2em}
$p p \rightarrow W^+ e^- W^- e^+$ 	&	$(3.447\pm0.87)\ex{-1}$ 	&	$(1.277\pm0.66)$\\
\rule{0pt}{1.2em}
$p p \rightarrow W^+ e^- W^- \mu^+$ 	&	$(7.06\pm0.15)\ex{-3}$ 		&	$(3.141\pm0.027)\ex{-2}$\\
\rule{0pt}{1.2em}
$p p \rightarrow W^+ e^+ W^- \mu^-$ 	&	$(6.99\pm0.16)\ex{-3}$	 	&	$(3.206\pm0.027)\ex{-2}$\\
\rule{0pt}{1.2em}
$p p \rightarrow W^+ e^- W^- \tau^+$ 	&	$(1.037\pm0.020)\ex{-2}$ 	&	$(4.293\pm0.036)\ex{-2}$\\
\rule{0pt}{1.2em}
$p p \rightarrow W^+ e^+ W^- \tau^-$ 	&	$(1.015\pm0.021)\ex{-2}$ 	&	$(4.411\pm0.036)\ex{-2}$\\
\rule{0pt}{1.2em}
$p p \rightarrow W^+ \mu^- W^- \mu^+$ 	&	$(3.74\pm0.10)\ex{-1}$	 	&	$(1.279\pm0.017)$\\
\rule{0pt}{1.2em}
$p p \rightarrow W^+ \mu^- W^- \tau^+$ 	&	$(2.913\pm0.048)\ex{-3}$ 	&	$(1.096\pm0.007)\ex{-1}$\\
\rule{0pt}{1.2em}
$p p \rightarrow W^+ \mu^+ W^- \tau^-$ 	&	$(2.990\pm0.042)\ex{-2}$ 	&	$(1.139\pm0.007)\ex{-1}$\\
\rule{0pt}{1.2em}
$p p \rightarrow W^+ \tau^- W^- \tau^+$ &	$(4.27\pm0.10)\ex{-1}$		&	$(1.606\pm0.017)$\\
\hline
\rule{0pt}{1.2em}
$p p \rightarrow W^+ e^- W^+ e^-$ 	&	$(1.112\pm0.013)\ex{-4}$ 	&	$(4.261\pm0.028)\ex{-4}$\\ 
\rule{0pt}{1.2em}
$p p \rightarrow W^+ e^- W^+ \mu^-$ 	&	$(1.537\pm0.023)\ex{-3}$ 	&	$(5.810\pm0.050)\ex{-3}$\\ 
\rule{0pt}{1.2em}
$p p \rightarrow W^+ e^- W^+ \tau^-$ 	&	$(4.721\pm0.055)\ex{-3}$ 	&	$(1.761\pm0.016)\ex{-2}$\\ 
\rule{0pt}{1.2em}
$p p \rightarrow W^+ \mu^- W^+ \mu^-$ 	&	$(4.099\pm0.052)\ex{-3}$ 	&	$(1.514\pm0.013)\ex{-2}$\\ 
\rule{0pt}{1.2em}
$p p \rightarrow W^+ \mu^- W^+ \tau^-$ 	&	$(2.704\pm0.036)\ex{-2}$ 	&	$(1.062\pm0.093)\ex{-1}$\\ 
\rule{0pt}{1.2em}
$p p \rightarrow W^+ \tau^- W^+ \tau^-$ &	$(4.614\pm0.065)\ex{-2}$ 	&	$(1.729\pm0.016)\ex{-1}$\\
\hline \hline
 \end{tabular}
\mycaption{Cross-sections for the processes with $W^+ \ell^- W^\pm \ell^\mp$ as final states (lepton number violating processes in lower section). A cut on the invariant lepton mass of 10 GeV has been assumed.}
\label{tab_results2}
\end{center}
\end{table}

In \Tab~\ref{tab_results2}, we give cross sections 
for the $2\to 4$ processes
containing two $W$-bosons. Note that we do not give the 
corresponding ones containing a $Z$-boson, see \equ{ZWll},
which are smaller because the corresponding contributions are those of
$2\to 3$ processes plus an additional $Z$-boson, attached to all 
internal and external lines in case of the lepton flavor mixing/violating
final states. As expected, the cross sections of lepton flavor conserving and lepton
flavor mixing final states are about two orders of magnitude
smaller than the ones of  the corresponding $2\to 3$ processes. However,
the cross sections for the lepton number violating processes are larger
than naively expected. This can be understood as follows:
in case of the $2\to 3$ processes all lepton number violating
contributions are due to the Majorana nature of the neutral fermions
and are suppressed by the pseudo-Dirac like nature of the heavy states.
In case of the $2\to 4$ processes, there are additional contributions
which are proportional to the momentum of the off-shell neutral particles
times two powers of lepton number violating couplings, e.g.\ they
are proportional
to $|c_i d_i|^2$. Performing
an approximate diagonalization of the neutral fermion mass matrix
 as done in Appendix \ref{app:approxdiag}, one sees that
this combination of couplings does not vanish in the limit of vanishing
neutrino masses as they are roughly proportional to
\begin{equation}
\frac{(a_i \kappa_1 + b_i \kappa_2)^4}{M^4_N m^4_\xi}
\end{equation}
e.g.\ they only vanish in the limit where either one of the heavy
masses goes to infinity or both couplings, $\kappa_1$ and $\kappa_2$,
to zero.
This is a consequence
of the fact that $\xi$ and $\xi'$ form a vector-like representation
of $SU(2)$. 

In summary we find that one should be able to detect the $SU(2)$ doublets
up to masses of about 1 TeV and show that they carry lepton number.
However, it turns out that the cross sections for the processes violating
total lepton number are on the edge to be discovered, as they would require
at least a luminosity of  
the order of $ab^{-1}$ in the most optimistic cases,
\eg, by considering at least 10 events without any background 
considerations due to detector effects.

\section{Summary and conclusions}
\label{sec:summary}

In this work we have studied neutrino mass generation from higher
than $d=5$ effective operators in supersymmetric models. While the
$d=5$ operator typically points towards the GUT scale, higher dimensional
operators may be generated by mediators observable at the LHC.
If any $d=5$ contribution is to be forbidden, a discrete symmetry is
needed, which can be used to control the dimension of the effective operator
generically dominating neutrino mass. While this discrete symmetry
is to be softly broken by the $\mu$-term of the MSSM, the $\mathbb Z_3$ symmetry,
the SUSY Lagrangian is invariant under, can be naturally used in the NMSSM. 
We have also taken into account  that in the NMSSM, higher than $d=5$ 
effective operators leading to neutrino mass may include the NMSSM
scalar and the two Higgs doublets. While the NMSSM scalar can be used
 in $d=6$ and $d=7$ effective operators, for $d>7$, only Higgs doublets
are allowed in the effective operators, since otherwise lower dimensional
effective operators are generated as well. Therefore, we have focused
on the $d=7$ operator $L L H_u H_u H_d H_u$ as the simplest possible example
in the following, which respects this line of argumentation.

For this operator, we have derived the list of possible decompositions
at tree level systematically. The results have been similar to an earlier
work~\cite{Bonnet:2009ej}, with the exception that some topologies
have been forbidden by the holomorphicity of the superpotential. Many of
the derived decompositions can be regarded as extensions of the usual
type~I,~II, or~III see-saw mechanisms because of a similar field
content. Models with two extra heavy fermion singlets, for example, lead to
inverse see-saw scenarios if the additional mediators are integrated out,
where the lepton number violating term is naturally suppressed by the
mediator mass. As a peculiarity of supersymmetry, we have identified that 
singlet scalars are potentially harmful because they may induce lower dimensional operators
dominating neutrino mass if similar to the NMSSM scalar. Therefore,
we have chosen an example with two extra fermion singlets and heavy lepton doublets
which are vector-like under $SU(2)$. We have also demonstrated how
the flavor structure for normal and inverted hierarchy can be easily
implemented using two generations of the heavy fermion singlets.

Focusing on the new fermions, we have demonstrated that parts of the model
can already be tested with the current LHC run at 7~TeV by
 displaced vertices, and at 14~TeV we expect that it can  be tested up to masses of several 
hundred GeV for the $SU(2)$ doublets.  We have also seen that the cross sections of 
some of the lepton number violating processes are larger than naively expected,
but still on the edge of observability at the LHC.

\section*{Acknowledgments}

We would like to thank F.~Bonnet and M.~Hirsch for useful discussions.
MBK acknowledges support from Research Training Group 1147
``Theoretical astrophysics and particle physics'' of Deutsche
Forschungsgemeinschaft. WW would like to acknowledge support from  
Deutsche Forschungsgemeinschaft, grants WI
2639/2-1 and WI 2639/3-1. W.P. is partially  supported by the German
Ministry of Education and Research (BMBF) under contract 05HT6WWA
 and by the Alexander von Humboldt Foundation. 

\begin{appendix}

\section{Possible decompositions for the operator $\boldsymbol{L L H_u H_u H_d H_u}$}
\label{app:uvcompl}

\begin{figure}[t]
\begin{center}
 \includegraphics[width=.9\linewidth]{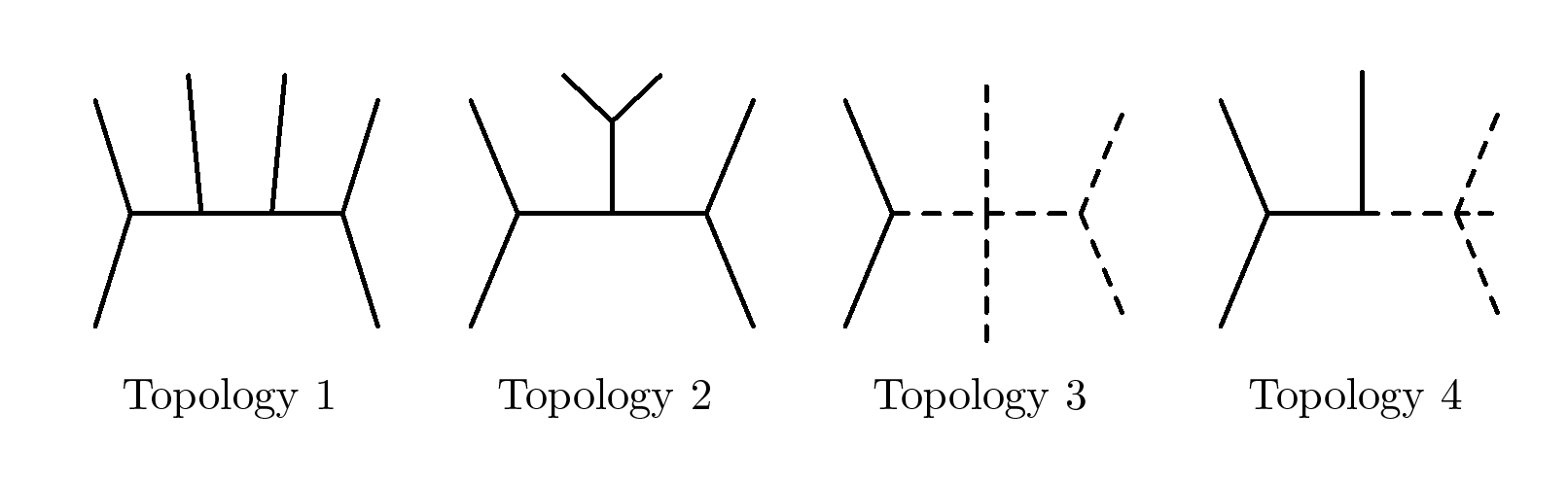}
 \end{center}
 \mycaption{\label{fig:topologies} Possible topologies for the effective $d=7$ operator $L L H_u H_u H_d H_u$. Topologies 3 and 4 cannot be realized in SUSY. Solid lines are either fermions or scalars, dashed lines are always scalars.}
\end{figure}

In this appendix, we systematically discuss underlying, more fundamental models leading to operator \#3 in  \Tab~\ref{tab:opOverviewMSSM}  or \Tab~\ref{tab:opOverviewNMSSM} at tree level. The results are similar to \Ref~\cite{Bonnet:2009ej} in the THDM.
The possible topologies for the decompositions can be found in \figu{topologies}. In SUSY, however, topologies~3 and~4 can be excluded. This is due to the fact that scalar couplings in SUSY have to be of the type $\phi^\dagger \phi$, $\phi^\dagger \phi \phi$, $\phi^\dagger \phi^\dagger \phi$ or $\phi^\dagger \phi^\dagger \phi \phi$, since they are generated by F-terms and D-terms, as a consequence of holomorphy. In topology~3, the scalar four-vertex has to be of the type $HHX^*X^*$, where $X$ is a heavy virtual scalar field. The three-vertex must be $HHX^*$. These two vertices can not be connected by a propagator $\Delta_X$. Hence topology 3 can not be realized in SUSY. In topology~4, the four-scalar vertex can only be of the type $H_d H_u H_u X^{(*)}$ or $H_u H_u H_u X^{(*)}$ in order to produce an effective operator of the type $LLH_uH_uH_dH_u$. The only possible scalar four couplings allowed by SUSY, however, are of the type $\phi^\dagger \phi^\dagger \phi \phi$. Hence also topology 4 is not possible.
\begin{table}[p]
\begin{center}
\hspace*{-15mm}
 \begin{tabular}{rccc}
\hline \hline
\# & Operator & Top. & Mediators
\\
\hline \rule[0em]{0em}{1.2em}  
1
&
     $(H_{u} {\rm i} \tau^{2} \overline{L^{c}})
     (H_{u} {\rm i} \tau^{2} L)(H_{d} {\rm i}
     \tau^{2} H_{u})$
&
2
&
${\bf 1}^{R}_{0}$, ${\bf 1}^{L}_{0}$, ${\bf 1}^{s}_{0}$
\\
2
&
     $(H_{u} {\rm i} \tau^{2} \vec{\tau} \overline{L^{c}})
     (H_{u} {\rm i} \tau^{2} L)
     (H_{d} {\rm i} \tau^{2} \vec{\tau} H_{u})$
&
2
&
${\bf 3}^{R}_{0}$, ${\bf 3}^{L}_{0}$
${\bf 1}^{R}_{0}$, ${\bf 1}^{L}_{0}$,
${\bf 3}^{s}_{0}$
\\
3
&
     $(H_{u} {\rm i} \tau^{2} \vec{\tau} \overline{L^{c}})
     (H_{u} {\rm i} \tau^{2} \vec{\tau} L)
     (H_{d} {\rm i} \tau^{2} H_{u})$
&
2
&
${\bf 3}^{R}_{0}$, ${\bf 3}^{L}_{0}$, ${\bf 1}^{s}_{0}$
\\
4
&
$ 
     (-{\rm i} \epsilon^{abc})
     (H_{u} {\rm i} \tau^{2} \tau^{a} \overline{L^{c}})
     (H_{u} {\rm i} \tau^{2} \tau^{b} L)
     (H_{d} {\rm i} \tau^{2} \tau^{c} H_{u})$
&
2
&
${\bf 3}^{R}_{0}$, ${\bf 3}^{L}_{0}$, ${\bf 3}^{s}_{0}$
\\
  	5 
&
     $(\overline{L^{c}} {\rm i} \tau^{2} \vec{\tau} L) 
     (H_{d} {\rm i} \tau^{2} H_{u}) 
     (H_{u} {\rm i} \tau^{2} \vec{\tau}  H_{u})$
&
	2
&
	${\mathbf 3}^{s}_{+1}$, ${\mathbf 3}^{s}_{+1}$, ${\mathbf 1}^{s}_{0}$ \\
  \hline \rule[0em]{0em}{1.2em}  
 	6 
&
     $(-\ii\epsilon_{abc})(\overline{L^{c}} {\rm i} \tau^{2} \tau_a L) 
     (H_{d} {\rm i} \tau^{2} \tau_b H_{u}) 
     (H_{u} {\rm i} \tau^{2} \tau_c  H_{u})$
&
	2
&
	${\mathbf 3}^{s}_{+1}$, ${\mathbf 3}^{s}_{+1}$, ${\mathbf 3}^{s}_{0}$ \\
´7
&
$(H_{u} {\rm i} \tau^{2} \overline{L^{c}})
(L {\rm i} \tau^{2} \vec{\tau} H_{d})
(H_{u} {\rm i} \tau^{2} \vec{\tau} H_{u})$
&
2
&
${\bf 1}^{R}_{0}$, ${\bf 1}^{L}_{0}$,
${\bf 3}^{R}_{-1}$, ${\bf 3}^{L}_{-1}$,
${\bf 3}^{s}_{+1}$
\\
8
&
$(-{\rm i} \epsilon^{abc})
(H_{u} {\rm i} \tau^{2} \tau^{a} \overline{L^{c}})
(L {\rm i} \tau^{2} \tau^{b} H_{d})
(H_{u} {\rm i} \tau^{2} \tau^{c} H_{u})$
&
2
&
${\bf 3}^{R}_{0}$, ${\bf 3}^{L}_{0}$,
${\bf 3}^{R}_{-1}$, ${\bf 3}^{L}_{-1}$,
${\bf 3}^{s}_{+1}$
\\
9
&
$(H_{u} {\rm i} \tau^{2} \overline{L^{c}})
 ({\rm i} \tau^{2} H_{u})
 (L)
 (H_{d} {\rm i} \tau^{2} H_{u})$
&
1
&
${\bf 1}^{R}_{0}$, ${\bf 1}^{L}_{0}$, 
${\bf 2}^{R}_{-1/2}$, ${\bf 2}^{L}_{-1/2}$,
${\bf 1}^{s}_{0}$
\\
10
&
$(H_{u} {\rm i} \tau^{2} \vec{\tau} \overline{L^{c}})
 ({\rm i} \tau^{2} \vec{\tau} H_{u})
 (L)
 (H_{d} {\rm i} \tau^{2} H_{u})$
&
1
&
${\bf 3}^{R}_{0}$, ${\bf 3}^{L}_{0}$, 
${\bf 2}^{R}_{-1/2}$, ${\bf 2}^{L}_{-1/2}$,
${\bf 1}^{s}_{0}$
\\
\hline \rule[0em]{0em}{1.2em}  
11
&
$(H_{u} {\rm i} \tau^{2} \overline{L^{c}})
 ({\rm i} \tau^{2} H_{u})
 (\vec{\tau} L)
 (H_{d} {\rm i} \tau^{2} \vec{\tau} H_{u})$
&
1
&
${\bf 1}^{R}_{0}$, ${\bf 1}^{L}_{0}$, 
${\bf 2}^{R}_{-1/2}$, ${\bf 2}^{L}_{-1/2}$,
${\bf 3}^{s}_{0}$
\\
12
&
$(H_{u} {\rm i} \tau^{2} \tau^{a} \overline{L^{c}})
 ({\rm i} \tau^{2} \tau^{a} H_{u})
 (\tau^{b} L)
 (H_{d} {\rm i} \tau^{2} \tau^{b} H_{u})$
&
1
&
${\bf 3}^{R}_{0}$, ${\bf 3}^{L}_{0}$, 
${\bf 2}^{R}_{-1/2}$, ${\bf 2}^{L}_{-1/2}$,
${\bf 3}^{s}_{0}$
\\
13
&
     $(H_{u} {\rm i} \tau^{2} \overline{L^{c}})(L) 
     ({\rm i} \tau^{2} H_{u})
     (H_{d} {\rm i} \tau^{2} H_{u})$
&
1
&
${\bf 1}^{R}_{0}$, ${\bf 1}^{L}_{0}$,
${\bf 2}^{s}_{+1/2}$, ${\bf 1}^{s}_{0}$
\\
14
&
     $(H_{u} {\rm i} \tau^{2} \vec{\tau} \overline{L^{c}})
     (\vec{\tau} L) 
     ({\rm i} \tau^{2} H_{u})
     (H_{d} {\rm i} \tau^{2} H_{u})$
&
1
&
${\bf 3}^{R}_{0}$, ${\bf 3}^{L}_{0}$,
${\bf 2}^{s}_{+1/2}$, ${\bf 1}^{s}_{0}$
\\
15
&
     $(H_{u} {\rm i} \tau^{2} \overline{L^{c}})(L) 
     ({\rm i} \tau^{2} \vec{\tau} H_{u})
     (H_{d} {\rm i} \tau^{2} \vec{\tau} H_{u})$
&
1
&
${\bf 1}^{R}_{0}$, ${\bf 1}^{L}_{0}$,
${\bf 2}^{s}_{+1/2}$, ${\bf 3}^{s}_{0}$
\\
\hline \rule[0em]{0em}{1.2em}  
16
&
     $(H_{u} {\rm i} \tau^{2} \tau^{a} \overline{L^{c}})( \tau^{a} L) 
     ({\rm i} \tau^{2} \tau^{b} H_{u})
     (H_{d} {\rm i} \tau^{2} \tau^{b} H_{u})$
&
1
&
${\bf 3}^{R}_{0}$, ${\bf 3}^{L}_{0}$,
${\bf 2}^{s}_{+1/2}$, ${\bf 3}^{s}_{0}$
\\
17
&
     $(H_{u} {\rm i} \tau^{2} \overline{L^{c}})
     (H_{d}) ({\rm i} \tau^{2} H_{u}) 
     (H_{u} {\rm i} \tau^{2} L)$
&
1
&
${\bf 1}^{R}_{0}$, ${\bf 1}^{L}_{0}$,
${\bf 2}^{R}_{-1/2}$, ${\bf 2}^{L}_{-1/2}$
\\
18
&
     $(H_{u} {\rm i} \tau^{2}\vec{\tau} \overline{L^{c}})
     (\vec{\tau} H_{d}) 
     ({\rm i} \tau^{2} H_{u}) 
     (H_{u} {\rm i} \tau^{2} L)$
&
1
&
${\bf 3}^{R}_{0}$, ${\bf 3}^{L}_{0}$,
${\bf 2}^{R}_{-1/2}$, ${\bf 2}^{L}_{-1/2}$,
${\bf 1}^{R}_{0}$, ${\bf 1}^{L}_{0}$
\\
19
&
     $(H_{u} {\rm i} \tau^{2} \overline{L^{c}})
     (H_{d}) ({\rm i} \tau^{2} \vec{\tau} H_{u}) 
     (H_{u} {\rm i} \tau^{2} \vec{\tau} L)$
&
1
&
${\bf 1}^{R}_{0}$, ${\bf 1}^{L}_{0}$,
${\bf 2}^{R}_{-1/2}$, ${\bf 2}^{L}_{-1/2}$,
${\bf 3}^{R}_{0}$, ${\bf 3}^{L}_{0}$
\\
20
&
     $(H_{u} {\rm i} \tau^{2} \tau^{a} \overline{L^{c}})
     (\tau^{a} H_{d}) ({\rm i} \tau^{2} \tau^{b} H_{u}) 
     (H_{u} {\rm i} \tau^{2} \tau^{b} L)$
&
1
&
${\bf 3}^{R}_{0}$, ${\bf 3}^{L}_{0}$,
${\bf 2}^{R}_{-1/2}$, ${\bf 2}^{L}_{-1/2}$,
\\
 \hline \rule[0em]{0em}{1.2em}  
  21
&
$(\overline{L^{c}} {\rm i} \tau^{2} \tau^{a} L) 
(H_{u} {\rm i} \tau^{2} \tau^{a})
(\tau^{b} H_{d})
(H_{u} {\rm i} \tau^{2} \tau^{b} H_{u})$
&
1
&
${\mathbf 3}^{s}_{+1}$, ${\mathbf 2}^{s}_{+1/2}$ , ${\mathbf 3}^{s}_{+1}$ \\
22
&
$(\overline{L^{c}} {\rm i} \tau^{2} \tau^{a} L) 
(H_{d} {\rm i} \tau^{2} \tau^{a})
(\tau^{b} H_{u})
(H_{u} {\rm i} \tau^{2} \tau^{b} H_{u})$
&
1
&
${\mathbf 3}^{s}_{+1}$, ${\mathbf 2}^{s}_{+3/2}$, ${\mathbf 3}^{s}_{+1}$ \\
23
&
$(\overline{L^{c}} {\rm i} \tau^{2} \vec{\tau} L) 
(H_{u} {\rm i} \tau^{2} \vec{\tau})
(H_{u})
(H_{d} {\rm i} \tau^{2} H_{u})$
&
1
&
${\mathbf 3}^{s}_{+1}$, ${\mathbf 2}^{s}_{+1/2}$, ${\mathbf 1}^{s}_{0}$\\
24
&
$(\overline{L^{c}} {\rm i} \tau^{2} \tau^{a} L) 
(H_{u} {\rm i} \tau^{2} \tau^{a})
(\tau^{b} H_{u})
(H_{d} {\rm i} \tau^{2} \tau^{b} H_{u})$
&
1
&
${\mathbf 3}^{s}_{+1}$, ${\mathbf 2}^{s}_{+1/2}$, ${\mathbf 1}^{s}_{0}$\\
25
&
$(H_{d} {\rm i} \tau^{2} H_{u}) 
     (\overline{L^{c}} {\rm i} \tau^{2})
     (\vec{\tau} L)
     (H_{u} {\rm i} \tau^{2} \vec{\tau} H_{u})$
&
1
&
${\bf 1}^{s}_{0}$, 
${\bf 2}^{L}_{+1/2}$, ${\bf 2}^{R}_{+1/2}$,
${\bf 3}^{s}_{+1}$
\\
\hline \rule[0em]{0em}{1.2em}  
26
&
$(H_{d} {\rm i} \tau^{2} \tau^{a} H_{u}) 
     (\overline{L^{c}} {\rm i} \tau^{2} \tau^{a})
     (\tau^{b} L)
     (H_{u} {\rm i} \tau^{2} \tau^{b} H_{u})$
&
1
&
${\bf 3}^{s}_{0}$, 
${\bf 2}^{L}_{+1/2}$, ${\bf 2}^{R}_{+1/2}$,
${\bf 3}^{s}_{+1}$
\\
27
&
$(H_{u} {\rm i} \tau^{2} \overline{L^{c}})
({\rm i} \tau^{2} H_{d})
(\vec{\tau} L)
(H_{u} {\rm i} \tau^{2} \vec{\tau} H_{u})$
&
1
&
${\bf 1}^{R}_{0}$, ${\bf 1}^{L}_{0}$,
${\bf 2}^{R}_{+1/2}$, ${\bf 2}^{L}_{+1/2}$,
${\bf 3}^{s}_{+1}$
\\
28
&
$(H_{u} {\rm i} \tau^{2} \tau^{a} \overline{L^{c}})
({\rm i} \tau^{2} \tau^{a} H_{d})
(\tau^{b} L)
(H_{u} {\rm i} \tau^{2} \tau^{b} H_{u})$
&
1
&
${\bf 3}^{R}_{0}$, ${\bf 3}^{L}_{0}$,
${\bf 2}^{R}_{+1/2}$, ${\bf 2}^{L}_{+1/2}$,
${\bf 3}^{s}_{+1}$
\\
29
&
$(H_{u} {\rm i} \tau^{2} \overline{L^{c}})
(L)
({\rm i} \tau^{2} \vec{\tau} H_{d})
(H_{u} {\rm i} \tau^{2} \vec{\tau} H_{u})$
&
1
&
${\bf 1}^{R}_{0}$, ${\bf 1}^{L}_{0}$,
${\bf 2}^{s}_{+1/2}$, 
${\bf 3}^{s}_{+1}$
\\
30
&
$(H_{u} {\rm i} \tau^{2} \tau^{a} \overline{L^{c}})
(\tau^{a} L)
({\rm i} \tau^{2} \tau^{b} H_{d})
(H_{u} {\rm i} \tau^{2} \tau^{b} H_{u})$
&
1
&
${\bf 3}^{R}_{0}$, ${\bf 3}^{L}_{0}$,
${\bf 2}^{s}_{+1/2}$,
${\bf 3}^{s}_{+1}$
\\
\hline \rule[0em]{0em}{1.2em}  
31
&
$(\overline{L^{c}} {\rm i} \tau^{2} \tau^{a} H_{d})
     ({\rm i} \tau^{2} \tau^{a} H_{u})
     (\tau^{b} L)
     (H_{u} {\rm i} \tau^{2} \tau^{b} H_{u})$
&
1
&
${\bf 3}^{L}_{+1}$, ${\bf 3}^{R}_{+1}$,
${\bf 2}^{L}_{+1/2}$, ${\bf 2}^{R}_{+1/2}$,
${\bf 3}^{s}_{+1}$
\\
32
&
$(\overline{L^{c}} {\rm i} \tau^{2} \tau^{a} H_{d})
     (\tau^{a} L)
     ({\rm i} \tau^{2} \tau^{b} H_{u})
     (H_{u} {\rm i} \tau^{2} \tau^{b} H_{u})$
&
1
&
${\bf 3}^{L}_{+1}$, ${\bf 3}^{R}_{+1}$,
${\bf 2}^{s}_{+3/2}$,
${\bf 3}^{s}_{+1}$
\\
33
&
$(\overline{L^{c}} {\rm i} \tau^{2} \vec{\tau} H_{d})
({\rm i} \tau^{2} \vec{\tau} H_{u})
(H_{u})
(H_{u} {\rm i} \tau^{2} L)$
&
1
&
${\bf 3}^{L}_{+1}$, ${\bf 3}^{R}_{+1}$,
${\bf 2}^{L}_{+1/2}$, ${\bf 2}^{R}_{+1/2}$,
${\bf 1}^{L}_{0}$, ${\bf 1}^{R}_{0}$
\\
34
&
$(\overline{L^{c}} {\rm i} \tau^{2} \tau^{a} H_{d})
({\rm i} \tau^{2} \tau^{a} H_{u})
(\tau^{b} H_{u})
(H_{u} {\rm i} \tau^{2} \tau^{b} L)$
&
1
&
${\bf 3}^{L}_{+1}$, ${\bf 3}^{R}_{+1}$,
${\bf 2}^{L}_{+1/2}$, ${\bf 2}^{R}_{+1/2}$,
${\bf 3}^{L}_{0}$, ${\bf 3}^{R}_{0}$
\\
\hline \hline
\end{tabular}
\mycaption{\label{tab:med} Decompositions of the $d=7$ operator $L L H_u H_u H_d H_u$ at tree level. }
\end{center}
\end{table}
The most economical extensions of the (N)MSSM may use the superpartners of the SM fields as mediators. However,  at least at tree level and with R-Parity conservation, this is not possible. As all external fields, $L$, $H_u$ and $H_d$, have $R = +1$, a mediator with $R=-1$ would cause vertices where R-Parity is violated. As a consequence, we have to introduce additional fields as mediators, and
we also obtain superpartners for them.
The possible decompositions of the $d=7$ operator $L L H_u H_u H_d H_u$ at tree level are shown in \Tab~\ref{tab:med}, where the brackets refer to the vertices with external fields for any given topology. If $\vec \tau$ appears, the fields couple to a triplet mediator; if not, they couple to a singlet. The mediators are denoted by  ${\bf X}^{\mathcal{L}}_Y$, where
\begin{itemize}
\item
 ${\bf X}$ denotes the SU(2) nature, \ie, singlet ${\bf 1}$, doublet ${\bf 2}$, or triplet ${\bf 3}$.
\item
     $\mathcal{L}$ refers to the Lorentz nature, \ie, scalar ($s$), vector
     ($v$),  left-handed ($L$) or right-handed ($R$) chiral fermion.
\item
 $Y$ refers to the hypercharge $Y=Q-I^W_3$.
\end{itemize}
Besides the fixed sign of the scalars' hypercharges and the forbidden topologies~3 and~4, the decompositions are similar to the THDM case in \Ref~\cite{Bonnet:2009ej}. Note that $R$ and $L$ indicate right- and left-handed fermions, respectively, where the right-handed ones can also be represented by left-handed Weyl spinors after charge conjugation.   All charged scalar fields must have an additional partner of opposite charge (not listed) to make a mass term possible in the superpotential.

\section{Approximate diagonalization of neutral fermion mass matrix}
\label{app:approxdiag}

In our model the complete mass matrix including the flavor
structure is given by
\begin{eqnarray}
\left(
\begin{array}{ccccccccc}
 0 & 0 & 0 & v_u Y_{N,11} & v_u Y_{N,12} & 0 & 0 & 0 & 0
   \\
 0 & 0 & 0 & v_u Y_{N,21} & v_u Y_{N,22} & 0 & 0 & 0 & 0
   \\
 0 & 0 & 0 & v_u Y_{N,31} & v_u Y_{N,32} & 0 & 0 & 0 & 0
   \\
 v_u Y_{N,11} & v_u Y_{N,21} & v_u Y_{N,31} &
   0 & 0 & M_N & 0 & 0 & 0 \\
 v_u Y_{N,12} & v_u Y_{N,22} & v_u Y_{N,32} &
   0 & 0 & 0 & M_N \rho  & 0 & 0 \\
 0 & 0 & 0 & M_N & 0 & 0 & 0 & -k_1 v_d & k_2 v_u
   \\
 0 & 0 & 0 & 0 & M_N \rho  & 0 & 0 & k_1 v_d & k_2
   v_u \\
 0 & 0 & 0 & 0 & 0 & -k_1 v_d & k_1 v_d & 0 & -m_\xi \\
 0 & 0 & 0 & 0 & 0 & k_2 v_u & k_2 v_u & -m_\xi &
   0
\end{array}
\right)
\end{eqnarray}
Using the fact, the the left-handed neutrinos are essentially
massless compared to the heavy states we can exploit the usual
seesaw formulas to obtain approximate formulas for the entries
responsible for the mixing of the light states with the heavy states.
The mass matrix of the heavy states is given by
\begin{equation}
M_H = \left(
\begin{array}{cccccc}
 0 & 0 & M_N & 0 & 0 & 0 \\
 0 & 0 & 0 & M_N \rho  & 0 & 0 \\
 M_N & 0 & 0 & 0 & -k_1 v_d & k_2 v_u \\
 0 & M_N \rho  & 0 & 0 & k_1 v_d & k_2 v_u \\
 0 & 0 & -k_1 v_d & k_1 v_d & 0 & -m_\xi \\
 0 & 0 & k_2 v_u & k_2 v_u & -m_\xi & 0
\end{array}
\right)
\end{equation}
Neglecting the elements proportional to $k_i$ ($i=1,2$) this 
matrix is diagonalized by
\begin{equation}
R_H = \left(
\begin{array}{cccccc}
 \frac{1}{\sqrt{2}} & 0 & \frac{1}{\sqrt{2}} & 0 & 0 & 0 \\
 0 & \frac{1}{\sqrt{2}} & 0 & \frac{1}{\sqrt{2}} & 0 & 0 \\
 \frac{1}{\sqrt{2}} & 0 & -\frac{1}{\sqrt{2}} & 0 & 0 & 0 \\
 0 & \frac{1}{\sqrt{2}} & 0 & -\frac{1}{\sqrt{2}} & 0 & 0 \\
 0 & 0 & 0 & 0 & \frac{1}{\sqrt{2}} & \frac{1}{\sqrt{2}} \\
 0 & 0 & 0 & 0 & \frac{1}{\sqrt{2}} & -\frac{1}{\sqrt{2}}
\end{array}
\right)
\end{equation}
The part of the mixing matrix connecting the heavy states with 
the light states is given by
\begin{eqnarray}
U' &=& m M^{-1}_H R_H \nonumber \\ &=&
\left(
\begin{array}{cccccc}
  D_1 Y_{N,11} & D_2 Y_{N,12} & D_2' Y_{N,11} &
    D_1' Y_{N,12} & \frac{v_u v_d (
    k_2' Y_{N,12}- k_1' \rho Y_{N,11} )  
   }{\sqrt{2} M_N m_\xi \rho } & \frac{v_u
   v_d (k_2' \rho  Y_{N,11}-k_1' Y_{N,12} )}{\sqrt{2} M_N
   m_\xi \rho } \\
D_1 Y_{N,21} & D_2 Y_{N,22} &  D_2' Y_{N,21} &
    D_1' Y_{N,22} & \frac{v_u v_d( k_2' 
   Y_{N,22}-k_1' \rho Y_{N,21})}{\sqrt{2} M_N m_\xi \rho } & \frac{v_u
   v_d (k_2' \rho  Y_{N,21}-k_1' Y_{N,22})}{\sqrt{2} M_N
   m_\xi \rho } \\
 D_1 Y_{N,31} & D_2 Y_{N,32} &  D_2' Y_{N,31} &
   D_1' Y_{N,32} & \frac{v_u v_d( k_2' 
   Y_{N,32}-k_1' \rho  Y_{N,31})}{\sqrt{2} M_N m_\xi \rho } & \frac{v_u
   v_d ( k_2' \rho  Y_{N,31}-k_1' Y_{N,32})}{\sqrt{2} M_N
   m_\xi \rho }
\end{array}
\right) \nonumber \\
\end{eqnarray}
with
\begin{eqnarray}
m &=& \left(
\begin{array}{cccccc}
 v_u Y_{N,11} & v_u Y_{N,12} & 0 & 0 & 0 & 0 \\
 v_u Y_{N,21} & v_u Y_{N,22} & 0 & 0 & 0 & 0 \\
 v_u Y_{N,31} & v_u Y_{N,32} & 0 & 0 & 0 & 0
\end{array}
\right) \\
D_1 &=& \frac{v_u(M_N m_\xi+2 k_1 k_2 v_d v_u)}
             {\sqrt{2} M_N^2 m_\xi}  \\
D_2 &=&  \frac{v_u( \rho M_N m_\xi-2 k_1 k_2 v_d v_u)}
             {\sqrt{2} \rho^2 M_N^2 m_\xi} \\
D_1' &=& -\frac{v_u(\rho M_N m_\xi +2 k_1 k_2 v_d v_u)}
             {\sqrt{2} \rho^2 M_N^2 m_\xi}  \\
D_2' &=&  -\frac{v_u(M_N m_\xi-2 k_1 k_2 v_d v_u)}
             {\sqrt{2} M_N^2 m_\xi} \\
k_1' &=&   k_1 -k_2 \tan\beta \\
k_2' &=&   k_1 +k_2 \tan\beta
\end{eqnarray}
Here we have the following correspondence to the couplings in
Section \ref{sec:lhc}, \equ{lag_nxiw}:
\begin{eqnarray}
c_i = U'_{i5} \,\,,\,\, d_i = (U'_{i6})^*
\end{eqnarray}
which are the dominating ones for the lepton number violating processes. 

\end{appendix}


\end{document}